\documentclass[journal]{IEEEtran}

\ifCLASSOPTIONcompsoc
  \usepackage[nocompress]{cite}
\else
  \usepackage{cite}
\fi

\ifCLASSINFOpdf
\else
\fi

\usepackage{graphicx}
\usepackage{multirow}
\usepackage[dvipsnames]{xcolor}
\usepackage[utf8]{inputenc}
\usepackage{fourier} 
\usepackage{array}
\usepackage{subfiles}
\usepackage{subcaption}
\usepackage{algorithmic}
\usepackage{listings}
\usepackage{footnote}
\usepackage{lipsum}
\usepackage{color}
\usepackage{hyperref}
\usepackage{makecell}
\hypersetup{
    colorlinks=true,
    linkcolor=blue,
    filecolor=magenta,      
    urlcolor=cyan,
    pdftitle={Overleaf Example},
    pdfpagemode=FullScreen,
    }
\usepackage{xcolor}
\usepackage[ruled,vlined]{algorithm2e}
\graphicspath{{figures/}}
\usepackage{amsmath}
\usepackage{amssymb}
\usepackage{booktabs}

\usepackage{dsfont}

\makeatletter
\def\algbackskip{\hskip-\ALG@thistlm}
\makeatother

\usepackage{xr}
\makeatletter
\newcommand*{\addFileDependency}[1]{
  \typeout{(#1)}
  \@addtofilelist{#1}
  \IfFileExists{#1}{}{\typeout{No file #1.}}
}
\makeatother

\newcommand*{\myexternaldocument}[1]{
    \externaldocument{#1}
    \addFileDependency{#1.tex}
    \addFileDependency{#1.aux}
}

\myexternaldocument{File2}

\begin{document}
%
\title{UncertaintyFuseNet: Robust Uncertainty-aware Hierarchical Feature Fusion Model with Ensemble Monte Carlo Dropout for COVID-19 Detection}
%
%
%
%

\author{Moloud~Abdar, Soorena Salari, Sina Qahremani,
    Hak-Keung Lam,~\IEEEmembership{Fellow,~IEEE}, Fakhri Karray*,~\IEEEmembership{Fellow,~IEEE}, Sadiq Hussain, Abbas Khosravi,~\IEEEmembership{Senior Member,~IEEE,} U. Rajendra Acharya,~\IEEEmembership{Senior Member,~IEEE,} Vladimir Makarenkov,~Saeid~Nahavandi,~\IEEEmembership{Fellow,~IEEE}
\IEEEcompsocitemizethanks{\IEEEcompsocthanksitem M. Abdar, A. Khosravi, and S. Nahavandi are with the Institute for Intelligent Systems Research and Innovation (IISRI), Deakin University, Australia. S. Nahavandi is also with the Harvard Paulson School of Engineering and Applied Sciences, Harvard University, Allston, MA 02134 USA (e-mails: m.abdar1987@gmail.com, mabdar@deakin.edu.au \& abbas.khosravi@deakin.edu.au \& saeid.nahavandi@deakin.edu.au).\protect
\IEEEcompsocthanksitem S. Salari and S. Qahremani are with the Department of Electrical Engineering, Sharif University of Technology, Tehran, Iran (e-mails: soorena.salari374@gmail.com \& sinaqahremani@gmail.com).\protect 
\IEEEcompsocthanksitem  H.K. Lam is with the Centre for Robotics Research, Department of Engineering, King’s College London, London, United Kingdom (e-mail: hak-keung.lam@kcl.ac.uk). \protect
\IEEEcompsocthanksitem  F. Karray is with the Centre for Pattern Analysis and Machine Intelligence, Department of Electrical and Computer Engineering, University of Waterloo, Waterloo, ON, Canada and the Department of Machine Learning, Mohamed bin Zayed University of Artificial Intelligence, Abu Dhabi, UAE (e-mail: karray@uwaterloo.ca).\protect
\IEEEcompsocthanksitem  S. Hussain is with the System Administrator, Dibrugarh University, Dibrugarh, India (e-mail: sadiq@dibru.ac.in). \protect
\IEEEcompsocthanksitem  U. R. Acharya is with the Department of Electronics and Computer Engineering, Ngee Ann Polytechnic, Clementi, Singapore (e-mail: aru@np.edu.sg).\protect
\IEEEcompsocthanksitem  V. Makarenkov is with the Department of Computer Science, University of Quebec in Montreal, Montreal (QC), Canada (e-mail: makarenkov.vladimir@uqam.ca). \protect
\IEEEcompsocthanksitem {*} {Corresponding author: Fakhri Karray, karray@uwaterloo.ca}}
}

\markboth{IEEE TRANSACTIONS ON }%
{Moloud Abdar\MakeLowercase{\textit{et al.}}:  UncertaintyFuseNet: Fused Uncertainty and Attention-based Deep Learning Model with Robustness Consideration for COVID-19 Classification}
%
\IEEEtitleabstractindextext{%
\begin{abstract} The COVID-19 (Coronavirus disease 2019) pandemic has become a major global threat to human health and well-being. Thus, the development of computer-aided detection (CAD) systems that are capable to accurately distinguish COVID-19 from other diseases using chest computed tomography (CT) and X-ray data is of immediate priority. Such automatic systems are usually based on traditional machine learning or deep learning methods. Differently from most of existing studies, which used either CT scan or X-ray images in COVID-19-case classification, we present a simple but efficient deep learning feature fusion model, called UncertaintyFuseNet, which is able to classify accurately large datasets of both of these types of images.  We argue that the uncertainty of the model's predictions should be taken into account in the learning process, even though most of existing studies have overlooked it. We quantify the prediction uncertainty in our feature fusion model using effective Ensemble MC Dropout (EMCD) technique. A comprehensive simulation study has been conducted to compare the results of our new model to the existing approaches, evaluating the performance of competing models in terms of Precision, Recall, F-Measure, Accuracy and ROC curves. The obtained results prove the efficiency of our model which provided the prediction accuracy of 99.08\% and 96.35\% for the considered CT scan and X-ray datasets, respectively. Moreover, our UncertaintyFuseNet model was generally robust to noise and performed well with previously unseen data. The source code of our implementation is freely available at: \url{https://github.com/moloud1987/UncertaintyFuseNet-for-COVID-19-Classification}.
\end{abstract}

}

\maketitle

\IEEEdisplaynontitleabstractindextext

%
\IEEEpeerreviewmaketitle

\ifCLASSOPTIONcompsoc
\IEEEraisesectionheading{\section{Introduction}\label{sec:introduction}}
\else
\section{Introduction}
\label{sec:introduction}
\fi
\IEEEPARstart{T}{he}
2019 coronavirus (COVID-19) has been spreading astonishingly fast across the globe since its emergence in Wuhan, China in December 2019~\cite{narin2020automatic}. Its exact origin is still unknown~\cite{mak2021,DOMINGO2021111785}. Overall, the COVID-19 pandemic has caused a consecutive series of catastrophic losses worldwide, infecting more than 287 million people and causing around 5.4 million deaths around the world up to the present. The rapid spread of COVID-19 is continuing to threaten human’s life and health with the emergence of novel variants such as Delta and Omicron. All of this makes COVID-19 not only an epidemiological disaster, but also a psychological and emotional one. The uncertainties and grappling with the loss of normalcy caused by this pandemic provoke severe anxiety, stress and sadness among people.\\
Easy respiratory transmission of the disease from person to person triggers swift spread of the pandemic. While many of the COVID-19 cases show milder symptoms, the symptoms of the remaining cases are unfortunately life-critical. The health-care systems in many countries seem to have arrived to the point of collapse as the number of cases has been increasing drastically due to the fast propagation of some of its variants. Regarding the COVID-19 diagnostic, the reverse transcription polymerase chain reaction (RT-PCR) is one of the gold standards for COVID-19 detection. However, RT-PCR has a low sensitivity. Hence, many COVID-19 cases will not be recognized by this test and thus the patients may not get the proper treatments. These unrecognized patients pose a threat to the healthy population due to highly infectious nature of the virus. Chest X-ray (CXR) and Computed Tomography (CT) have been widely used to identify prominent pneumonia patterns in chest. These imaging technologies accompanied by artificial intelligence tools may be used to diagnose COVID-19 patients in a more accurate, fast and cost-effective manner. Failure to provide prompt detection and treatment of COVID-19 patients increases the mortality rate. Hence, the detection of  COVID-19 cases using deep learning models using both CXR and CT images may have huge potential in healthcare applications.  \\
In recent years, deep learning models have had the widespread applicability not only in medical imaging field but also in many other areas~\cite{wang2020recent,pourpanah2020review,luo2021dual, shamsi2021uncertainty}. These models have also been extensively applied for COVID-19 detection. It is critical to discriminate COVID-19 from other forms of pneumonia and flu. Farooq et al.~\cite{farooq2020covid} introduced an open-access dataset and the open-source code of their implementation using a CNN framework for distinguishing COVID-19 from analogous pneumonia cohorts from chest X-ray images. The authors designed their COVIDResNet model by utilizing a pre-trained ResNet-50 framework allowing them to improve the model's performance and reduce its training time. An automatic and accurate identification of COVID-19 using CT images helps radiologists to  screen patients in a better way. Zheng et al. in~\cite{zheng2020deep} proposed a fully automated system for COVID-19 detection from chest CT images. Their deep learning model, called COVNet, investigates visual features of the chest CT images. Moreover, Hall et al.~\cite{hall2020finding} presented a new deep learning model, named COVIDX-Net, to aid radiologists with COVID-19 detection  from CXR image data. The authors explored seven deep learning architectures, including DenseNet, VGG-19 and MobileNet v2.0. In another study, Abbas et al.~\cite{abbas2020classification} designed the Decompose, Transfer, and Compose (DeTraC) model of COVID-19 image classification using CXR data. A class decomposition approach was employed to identify irregularities in iCXR data by scrutinizing the class boundaries.\\
Segmentation also plays a key role in COVID-19 quantification applied to CT scan data. Chen et al.~\cite{chen2020residual} proposed a novel deep learning method for segmentation of COVID-19 infection regions automatically. Aggregated Residual Transformations were employed to learn a robust and expressive feature representation and the soft attention technique was applied to improve the potential of the system to distinguish several symptoms of COVID-19. However, we noticed that there are still some open issues in the recently proposed traditional machine learning and deep learning models for COVID-19 detection. For this reason, optimizing the existing models should be a priority in COVID-19 detection and classification. Ensemble and fusion-based models~\cite{ali2020smart} have shown outstanding performance in different medical applications. In the following, we provide more information about fusion-based models, discussing how they can be used in the framework of the deep learning approach. 
\subsection{Information fusion} 
Information fusion is initiated from data fusion. It can also be termed as multi-sensor information fusion~\cite{shanshanoriginal}, feature fusion for combining different features~\cite{xie2018fusing}, various biological sources~\cite{martinez2017machine, chen2020ai}, medical signals~\cite{yao2020multi}, or medical image fusion~\cite{james2014medical, acharya2016integrated, he2020feasibility}. Different data fusion models have been widely used in military applications. Their purpose was to integrate or correlate data of several sensors of different, or the same, type(s) to achieve better results than those yielded by a single sensor. Gradually, data fusion models have been converted into information fusion models. Information fusion does not rely on multi-sensor data only. Its areas of research and application have been growing drastically. The rapid emergence of network technologies allowed the information fusion to change from centralized single node information fusion to distributed information fusion. \\
Modern medicine nowadays depends on amalgamation of data and information from manifold sources that include structured imaging data, laboratory data, unstructured narrative data, and even observational or audio data in some cases~\cite{leslie2000influence}. Substantial clinical context is required for medical image interpretation to facilitate diagnostic decisions~\cite{boonn2009radiologist}. Imaging data are not only limited to radiology but also concern many other image-based medical specialties such as dermatology, ophthalmology, and pathology~\cite{jonas2020high}. Unstructured and structured clinical data from the electronic heath records (EHR) are crucial for clinically relevant medical image interpretation~\cite{kumar2020deep}. Clinically relevant models rely on automated diagnosis and classification systems that use both clinical data from EHR and medical imaging data. In various applications, such as video classification, autonomous driving, and medical data analysis, multimodal learning models use various imaging data along with other data types (data fusion approach). The current medical imaging paradigm showcases a drift where both pixel and EHR data are employed in fusion-domains for tackling complicated tasks which cannot be resolved by single modality. A wide variety of fusion techniques have been applied with traditional machine learning and deep learning techniques. This facilitates an increasing interest in several areas, each of which has its specific prerequisites. In medicine, customized predictions carry significant meaning as incorrect decisions are associated with severe costs due to associated ethical concerns and risk to human life~\cite{dusenberry2020analyzing}. Deep neural networks (DNNs) are now prevailing in many medical applications. The performance of DNNs can depend on either one DNN model or an ensemble of several DNN models, focusing on enhancing the accuracy of probabilistic predictions. Model's uncertainty is inherent in fitting DNNs, which is not well addressed in the literature, while some DNN models can use probabilities to capture data uncertainty. For example, when the mortality of the patients is predicted using intensive care unit (ICU) data, the state-of-the-art (SOTA) methods may be able to yield high values of the AUC-ROC statistics. However, these methods are unable to discriminate between the cases in which the model is certain about its predictions or fairly uncertain about them. Hence, there is an urgent need for examining the use of both model and data uncertainty, specifically in the context of predictive medicine. Recently proposed model uncertainty techniques include: function priors, deep ensembles, Monte Carlo (MC) dropout, and reparameterization-based variational Bayesian neural networks (BNNs). Thus, several clinical care problems can be efficiently addressed by DNNs integrating model uncertainty techniques.
\subsection{Uncertainty quantification (UQ)}
Many traditional machine learning and deep learning models have been developed not only for analysis of CXR and CT image data but also for many other medical applications, often yielding high accuracy results even for a limited number of images~\cite{shamsi2021uncertainty}. However, DNNs require a large number of data to fine-tune trainable parameters. A limited number of images usually leads to epistemic uncertainty. Trust is an issue for these models, deployed with lower numbers of training samples. Out-of-distribution (OoD) samples and discrimination between the training and testing samples make such models fail in real world applications. Lack of confidence in unknown or new cases is usually not reported for these models. However, this information is essential for the development of reliable medical diagnostic tools. These unknown samples, which are generally hard to predict, often have important practical value. It is essential to estimate uncertainties with an extra insight to their point estimates. This additional vision aims at enhancing the overall trustworthiness of the systems, allowing clinicians to know where they can trust predictions made by the models. The flawed decisions made by some models can be fatal for the patients at risk. Hence, proper uncertainty estimations are necessary to improve the efficiency of ML models making them trustworthy and reliable~\cite{abdar2020review, abdar2021uncertainty}. Trustworthy uncertainty estimates can facilitate clinical decision making, and more importantly, provide clinicians with appropriate feedback on the reliability of the obtained results. As discussed above, COVID-19 has had many negative effects on all aspects of human life around the world. The COVID-19 pandemic has caused millions of deaths worldwide. In this regard, our study attempts to propose a simple and accurate deep learning model, called UncertaintyFuseNet, for detecting COVID-19 cases. Our model includes an uncertainty quantification method to increase the reliability of the obtained results.

\subsection{Research Gaps}
\label{Sec:sec:RG}
Our comprehensive literature review helped us to identify several important research gaps related to the use of the COVID-19 detection/segmentation methods. Below, we list the most important of them:
\begin{itemize}
  \item There are no sufficient COVID-19 image data to develop accurate and robust deep learning models. This lack of data can impact the performance of deep learning approaches. 
  \item To the best of our knowledge, there are very few studies that have used both types of images (CT scan and X-ray) simultaneously. 
  \item There are very few studies that have examined the uncertainty of the COVID-19 predictions provided by deep learning models.  
  \item Moreover, we found that there are very few COVID-19 classification studies studies considering the model's robustness and its ability to process unknown data.
  \item The impressive effect of different feature fusion methods have received less attention in the COVID-19 classification research. It is worth noting that feature fusion techniques are very effective both for improving the model's performance and for dealing with uncertainty within ML and DL models.
\end{itemize}
\subsection{Main Contributions}
\label{Sec:sec:LR}
The main contributions of this study are as follows:
\begin{itemize}
  \item We proposed a novel feature fusion model for accurate detection of COVID-19 cases.
  \item We quantified the uncertainty in our proposed feature fusion model using effective Ensemble MC Dropout (EMCD) technique.
   \item The proposed feature fusion model demonstrates strong robustness to data contamination (data noise).
   \item Our new model provided very encouraging results in terms of unknown data detection.
\end{itemize}
The rest of this study is organized as follows. Section \ref{Sec:LR} summarizes a few relevant studies. Section \ref{Sec:PM} formulates the proposed methodology. The main experiments of this study are discussed in Section \ref{Sec:ER}. Section \ref{Sec:DI} presents the obtained results and provides a comprehensive comparison with existing studies. Finally, the conclusions are presented in Section \ref{Sec:Co}. 

The main characteristics of the proposed UncertaintyFuseNet model are as follows: (i) It is an accurate model with promising performance, (ii) It can be used efficiently to carry out classification analysis of large CT and X-ray image datasets, (iii) It quantifies the prediction uncertainty, (iv) It is a reliable model in terms of processing noisy data, and finally, (v) It allows for an accurate detection of out-of-distribution (OOD) samples. 
\section{Literature Review}
\label{Sec:LR}
In this section, we will briefly review a few recent studies conducted on COVID-19 detection/segmentation as well as those using UQ in medical image analysis. 

\subsection{COVID-19 Classification/Segmentation}
\label{Sec:sec:COVID}
It is crucial to recognize COVID-19 cases quickly to better manage and prevent the pandemic from further spreading. A wide variety of traditional machine learning and deep learning models have been used for for this purpose~\cite{abualigah2021novel}. For example, Pathak et al.~\cite{pathak2020deep} showed that deep transfer learning is a useful approach for COVID-19 classification. In another work, Ardakani et al.~\cite{ardakani2020application} analyzed 108 COVID-19 patients, those with viral pneumonia, and other atypical patients, using CT scan images. They tested ten CNN models to discriminate the COVID-19 group of patients from the non-COVID-19 cohorts. The Xception and ResNet-101 models demonstrated a superior performance for their data with an AUC value of 0.994 for both of them. Deep learning models can assist the clinicians and radiologists utilizing CXR scans for the detection of COVID-19. In this context, Khan et al.~\cite{khan2020coronet} introduced CoroNet, a deep CNN model, allowing for automated detection of COVID-19 cases. Xception architecture was used for pretraining and two publicly available X-ray datasets were used for classification of normal, pneumonia and COVID-19 cases. The model by Khan et al. yielded the accuracy of 95\% for 3-class (Normal vs Pneumonia vs COVID-19) classification.  In addition, their model demonstrated an overall accuracy of 89.6\% for 4-class (Pneumonia bacterial vs Pneumonia viral vs Normal vs COVID-19) classification. CoroNet provided promising results with minimal preprocessing of data. Chimmula et al.~\cite{chimmula2020time} used modern deep learning methods to design a COVID-19 prediction model (the Long Short-Term Memory (LSTM) method) using publicly available Canadian health authority and John Hopkins University data. The authors also scrutinized some vital features to predict probable stopping time and eventual trends of the pandemic. \\
Afshar et al.~\cite{afshar2020covid} devised a COVID-19 prediction approach based on Capsule networks (COVID-CAPS) and produced efficient results with smaller X-ray datasets. Their framework exhibited better performance than the existing CNN-based models. COVID-CAPS exhibited the AUC value of 0.97, the specificity value of 95.8\%, the value of sensitivity of 90\%, and the accuracy value of 95.7\% while dealing with a lower number of network parameters than its counterparts. Transfer learning and pretraining were used to further enhance the diagnostic nature of the framework and tested with a new X-ray dataset. The use of artificial intelligence (AI) to analyze CXR images for accurate COVID-19 patient triage is of supreme importance. Lack of systematic collection of CXR data for training of deep learning strategies hinders the proper diagnosis. To address this issue, Oh et al.~\cite{oh2020deep} presented a patch-based CNN technique for COVID-19 patient detection using a low number of trainable parameters. Punn et al.~\cite{punn2020automated} proposed the weighted class loss function and random oversampling methods for transfer learning for different SOTA deep learning models. They used posteroanterior CXR images for multiclass classification: (Pneumonia, COVID-19, and Normal cases) and for binary classification (COVID-19 and Normal cases). The experimental results of Punn et al. demonstrated that each of the models they considered was scenario-dependent, and that NASNetLarge showed better scores compared to its counterparts.
\subsection{Uncertainty Quantification in Medical Image Analysis}
\label{Sec:sec:LR}
There are numerous studies conducted on Uncertainty Quantification (UQ) in medical image analysis using traditional ML and DL methods. In this section, we briefly discuss a few recent studies which applied different UQ methods in medical image analysis. Deep CNNs don't facilitate uncertainty estimation in medical image segmentation, e.g., image-based (aleatoric) and model-based (epistemic) uncertainties, despite delivering the SOTA performance. Wang et al.~\cite{wang2019aleatoric} examined different types of uncertainties related to 3D and 2D medical image segmentation tasks at both structural and pixel levels. Moreover, they introduced test-time augmentation-based aleatoric uncertainty to measure the effect of various transformations of the input image on the output segmentation. MC simulation with prior distributions of parameters was used to estimate a distribution of predictions in an image acquisition model with noise and image transformations. It helped to formulate test-time augmentation.\\
The direct ventricle function index estimation and bi-ventricle segmentation can be used to tackle ventricle quantification issue. Luo et al.~\cite{luo2020commensal} introduced a unified bi-ventricle quantification approach based on commensal correlation between the direct area estimation and bi-ventricle segmentation. The authors devised a new deep commensal network (DCN) to combine these two commensal tasks into a unified framework based on the proposed commensal correlation loss. The proposed DCN ensured fast convergence, carrying out end-to-end optimization as well as uncertainty estimation with one-time inference. Colorectal cancer is one of the prime reasons of cancer-related fatalities around the globe. Its key precursors are colorectal polyps. Some modern CNNs based decision support systems for segmentation and detection of colorectal polyps provide an excellent performance. In another study, Ghoshal et al.~\cite{ghoshal2020estimating} used drop-weights based Bayesian CNN (BCNN) to measure uncertainty in deep learning methods to enhance the diagnostic performance. They demonstrated that accuracy of the prediction was highly correlated with the uncertainty in prediction. Recently, Mazoure et al.~\cite{mazoure2021} have presented the DUNEScan (Deep Uncertainty Estimation for Skin Cancer) web application performing in-depth analysis of uncertainty within some modern CNN models. DUNEScan relies on efficient Grad-CAM and UMAP methods to visualize the classification manifold for the user’s input, yielding key information about its closeness to skin lesion images available in the ISIC data repository.

\section{Proposed Methodology}
\label{Sec:PM}
This section includes two main sub-sections describing: (i) Basic deep learning models in sub-section \ref{Sec:sec:Base}, (ii) and our novel feature fusion model, UncertaintyFuseNet, in sub-section \ref{Sec:sec:PFM}. It may be noted that we also applied two traditional machine learning algorithms (\emph{i.e.}, $Random \ Forest$ (RF) and $Decision\ Tree$ (DT, max-depth=50 and n-estimators=200)) and compared their performances with the considered deep learning models.
\subsection{Basic Deep Learning Models}
\label{Sec:sec:Base}
\begin{figure}[h]
\centering
            \includegraphics[width=0.5\textwidth]{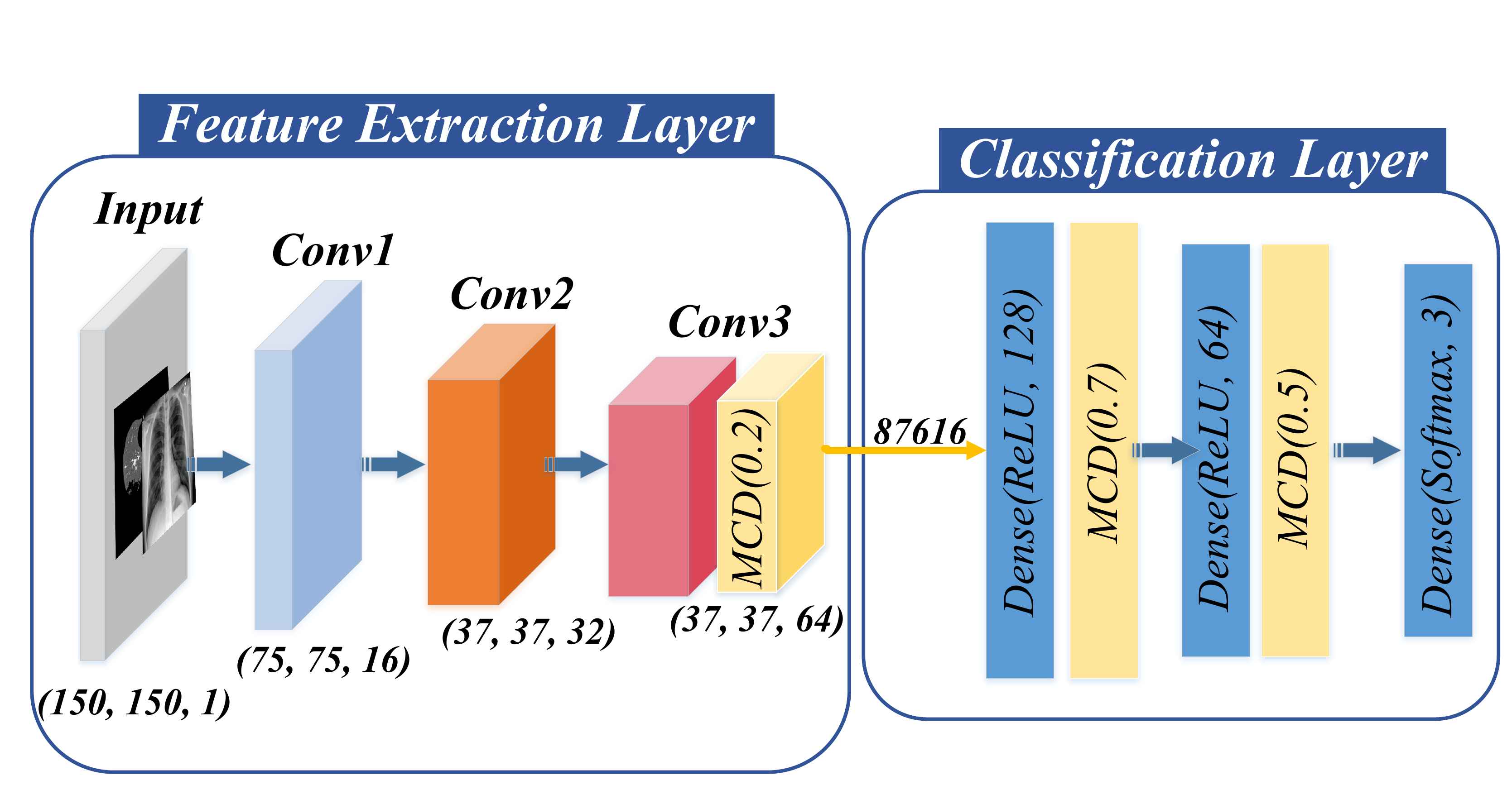}
    \caption{A general overview of the applied deep learning model Deep 1 (Simple CNN).}\label{SimpelCNN}
\end{figure}

\begin{figure}[h]
\centering
            \includegraphics[width=0.50\textwidth]{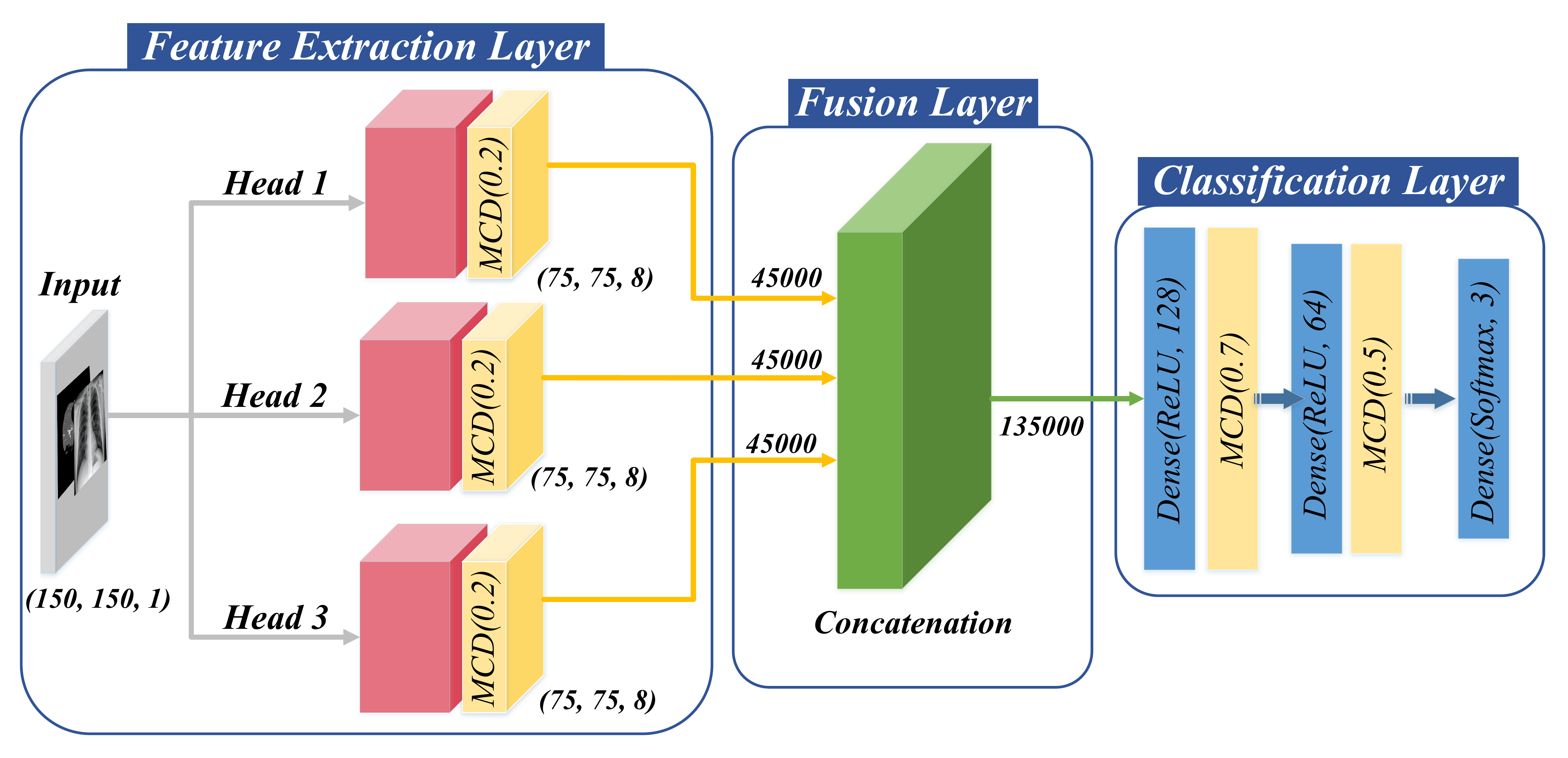}
    \caption{A general overview of the applied deep learning mode Deep 2 (Multi-headed CNN).}\label{Multi-headed}
\end{figure}

In this sub-section, we provide more details regarding two basic deep learning models: (i) Deep 1 (Simple CNN), and (ii) deep 2 (Multi-headed CNN). Figs. \ref{SimpelCNN} and \ref{Multi-headed} show deep 1 (Simple CNN) and deep 2 (Multi-headed CNN) models, respectively. 
The first deep learning model (Simple CNN) includes three convolutional layers followed by MC dropout in the feature extraction layer. The extracted features are then given to the classification layer, including three dense layers and MC dropout. More details of the deep 1 model can be found in Fig. \ref{SimpelCNN}. In our second deep learning model, deep 2, \emph{i.e.}, multi-headed CNN, comprises three main heads (as feature extractors). The extracted features in each branch are then given to the fusion layers, followed by the classification layer as illustrated in Fig. \ref{Multi-headed}. 

\subsection{Proposed Feature Fusion Model: UncertaintyFuseNet}
\label{Sec:sec:PFM}
Feature fusion is an approach used to combine features (different information) of the same sample (input) extracted by various methods. 
Assume $\Omega = \{\xi\mid\xi \in \mathbb{R}^N\}$ be a training sample (image) space of $m$ labeled samples (images). 
Given $A = \{x\mid x \in \mathbb{R}^p\}$, $B = \{y\mid y \in \mathbb{R}^q\}$, ..., and $Z = \{n\mid n \in \mathbb{R}^k\}$, where $x$, $y$, ..., $n$ are the feature vectors of the same input sample $\xi$  extracted by various deep learning models, respectively. Therefore, the total feature fusion vector space $D^{ff}_{vs}$ obtained from different sources can be calculated as follows:

\begin{align}
\label{eq:FEFU}
D^{ff}_{vs} = Concatenate[A, B,..., Z],
\end{align}

In this study, after preprocessing the data, we feed our dataset to the model. Our model consists of two major branches: The first branch has five convolutional blocks. Each block is made up of two tandem convolutional layers followed by batch normalization and max-pooling layers. Also, the fourth and fifth blocks have dropout layers in their outputs. The second branch is a VGG16 transfer learning network whose output is used in the fusion layer. After two branches, the model is followed by a fusion layer that concatenates the third, fourth, and fifth convolutional layers' outputs with VGG16's output.

Finally, we used fully connected layers to process the fused features and classify the data. In this part, we have used four dense layers with 512, 128, 64, and 3 neurons with the ReLU activation function, respectively. The output of the first three dense layers has a dropout in their outcomes with a rate equal to 0.7, 0.5, and 0.3, respectively. \\
The stated model is not simplistic. Indeed, to boost the model's power in dealing with data and extracting high-quality features, we have employed a novel feature fusion approach combining different sources:
\begin{itemize}
\item We selected the third convolutional block's output as a fusion source to have a holistic perspective about the data distribution. These features help the model to consider the unprocessed and raw information and use it in the prediction.\\
\item We included the final and penultimate convolutional blocks' outputs in the feature fusion layer to have more accurate information. This feature gives a detailed view of the dataset to model and helps the model to process advanced classification features.\\
\item As has been suggested by recent pneumonia detection studies, where the pretrained networks have been successively used to create high-quality generalizable features, we used the output of VGG16 in the fusion layer.\\
\end{itemize}

\begin{figure*}[h]
\centering
            \includegraphics[width=0.64\textwidth]{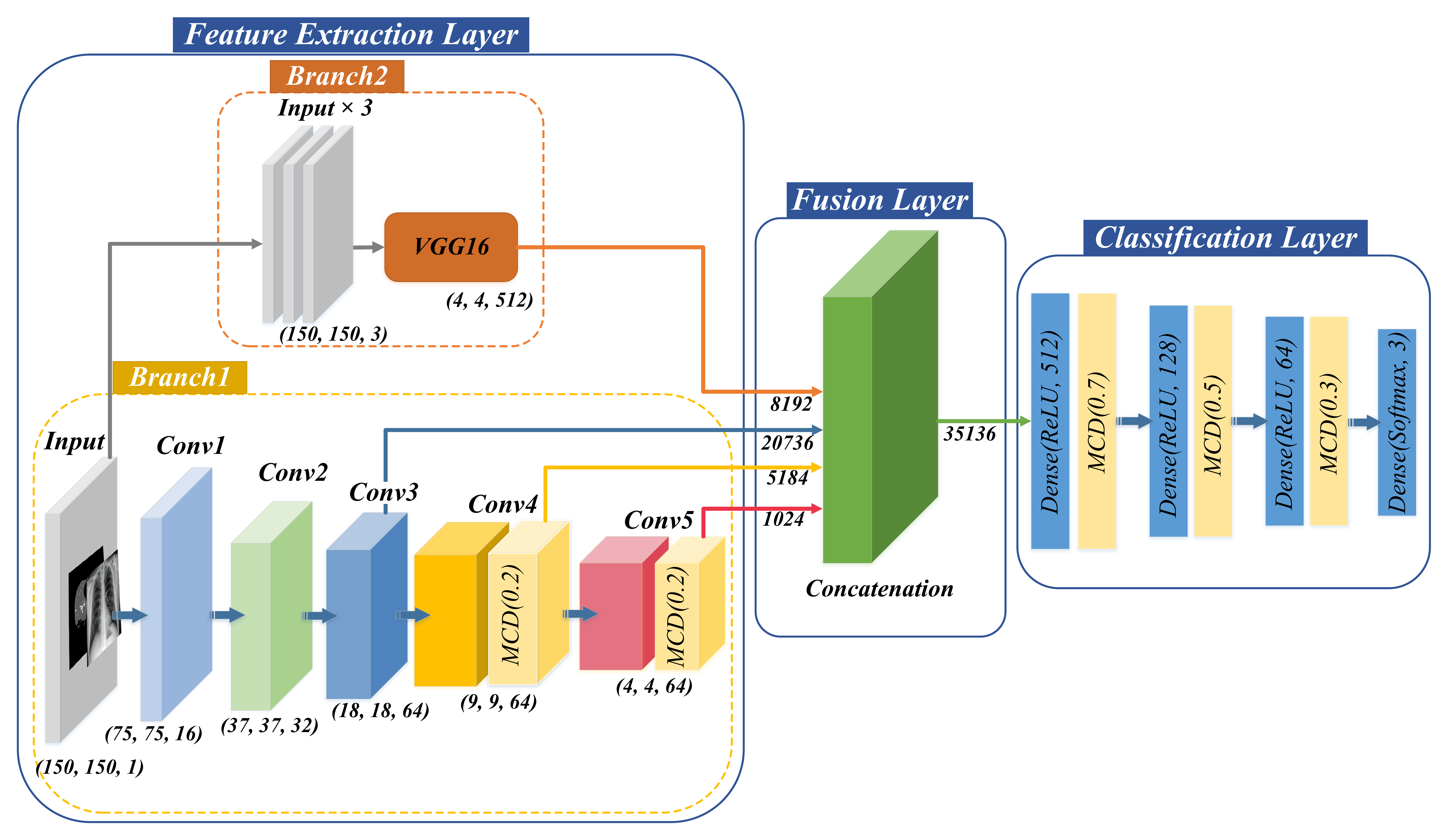}
    \caption{A general overview of the proposed UncertaintyFuseNet model inspired by a hierarchical feature fusion approach and EMCD.}\label{UncertaintyFuseNet}
\end{figure*}

The pseudo-code of the proposed UncertaintyFuseNet model for detecting the COVID-19 cases is reported in \textbf{Algorithm \ref{euclid}}. Its general view is illustrated in Fig. \ref{UncertaintyFuseNet}.\\

\begin{algorithm}[tb!]
\small
\caption{Pseudo-code of the proposed feature fusion model (UncertaintyFuseNet)}\label{euclid}
\KwIn{A gray-scale CT scan and X-ray image.\\
}
\KwOut{COVID-19 classification with higher certainty.\\
}
\textbf{Feature Extraction Layer: Branch1:}\\
$\textit{Conv1} \gets \textit{First Convolutional Block} \gets \textit{Input Image}$\\
$\textit{Conv2} \gets \textit{Second Convolutional Block} \gets \textit{Conv1}$\\
$\textit{Conv3} \gets \textit{Third Convolutional Block} \gets \textit{Conv2}$\\
$\textit{Conv4} \gets \textit{MCDropout(rate=0.2)} \gets \textit{Fourth Convolutional Block} \gets \textit{Conv3}$\\
 $\textit{Conv5} \gets \textit{MCDropout(rate=0.2)} \gets \textit{Fifth Convolutional Block} \gets \textit{Conv4}$ \textbf{Branch2:}\\
  $\textit{VGG features} \gets \textit{VGG16 Block} \gets \textit{Input Image}$ \\
\textbf{Fusion Layer}:\\ 
 $\textit{X} \gets \text{Concatenation of }\textit{(Conv3, Conv4, Conv5, VGG features)}$\\
\emph{\textbf {{\normalfont\textbf{{Classification Layer}}}}}: 
   $\textit{X} \gets \textit{Dense(X, units=512, activation=ReLU)}$
   $\textit{X} \gets \textit{MCDropout(X, rate=0.7)}$
   $\textit{X} \gets \textit{Dense(X, units=128, activation=ReLU)}$
  $\textit{X} \gets \textit{MCDropout(X, rate=0.5)}$
   $\textit{X} \gets \textit{Dense(X, units=64, activation=ReLU)}$
   $\textit{X} \gets \textit{MCDropout(X, rate=0.3)}$
   $\textit{Output} \gets \textit{Dense(X, units=3, activation=softmax)}$\\

\end{algorithm}

It should be noted that the detailed information about Convolution blocks in Figs. \ref{SimpelCNN}, \ref{Multi-headed}, and \ref{UncertaintyFuseNet} is reported in Table \ref{DETTAA}, in the Supplementary Material. To generate the final prediction, after training the applied models with uncertainty module, we have first run each model $N$ times. Thereafter, we have averaged the predicted softmax probabilities (outputs) in those $N$ random predictions of data $x$ through UncertaintyFuseNet and stochastic sampling dropout mask $\mathbf{w}_{t}$ for each single prediction.
\\
\begin{align}
\label{eq:SOF1}
\hat{\mathbf{y}}_{t} =  \operatorname{Softmax}\left(\mathbf{\textit{UncertaintyFuseNet}}\left(\mathbf{x}; \mathbf{w}_{t}\right)\right),
\end{align}

\begin{align}
\label{eq:SOF2}
\hat{\mathbf{y}}_{*}=\frac{1}{N} \sum_{0}^{N} \hat{\mathbf{y}}_{t}.
\end{align}
We then used the model ensembling and acquired predictions from the $N$ trained models with various weight distributions and initialized weights using this strategy. This allowed us to improve the model's performance drastically. \\

\begin{align}
\label{eq:SOF3}
Predicted\textunderscore Class =\operatorname{Argmax}(\hat{\mathbf{y}}_{*}).
\end{align}

The pseudo-code of the applied EMCD procedure included in our UncertaintyFuseNet model for detecting COVID-19 cases is reported in \textbf{Algorithm \ref{euclid20}}.\\

\begin{algorithm}[tb!]
\small
\caption{Ensemble MC Dropout (EMCD) procedure}\label{euclid20}
\textit{Predictions} = 0\\
\For{$k = 1$, $k{+}{+}$, \normalfont{while} $k < i$}
    {
    $\textit{Probability} \gets \textit{UncertaintyFuseNet} \gets \textit{Input Image}$\\
    $\textit{Predictions} \gets \textit{Predictions + Probability}$\\
    }
 $\textit{Predictions} \gets \textit{Mean(Predictions)}$\\ 
$\textit{Predicted Class} \gets \textit{Argmax(Predictions)}$
\end{algorithm}


\section{Experiments}
\label{Sec:ER}
In this section, we present : the data considered in our study (see sub-section \ref{Sec:sec:DA}), the results obtained using our new model (see sub-section \ref{Sec:sec:ER}), the results showing that our new model is robust against noise (see sub-section\ref{Sec:sec:RAN}), and the results showing how our new model copes with unknown data (see sub-section \ref{Sec:sec:UNKN}).
\subsection{Datas considered}
\label{Sec:sec:DA}
\begin{figure}[h]
\centering
    \begin{subfigure}[b]{0.18\textwidth}
            \centering
            \includegraphics[width=\textwidth]{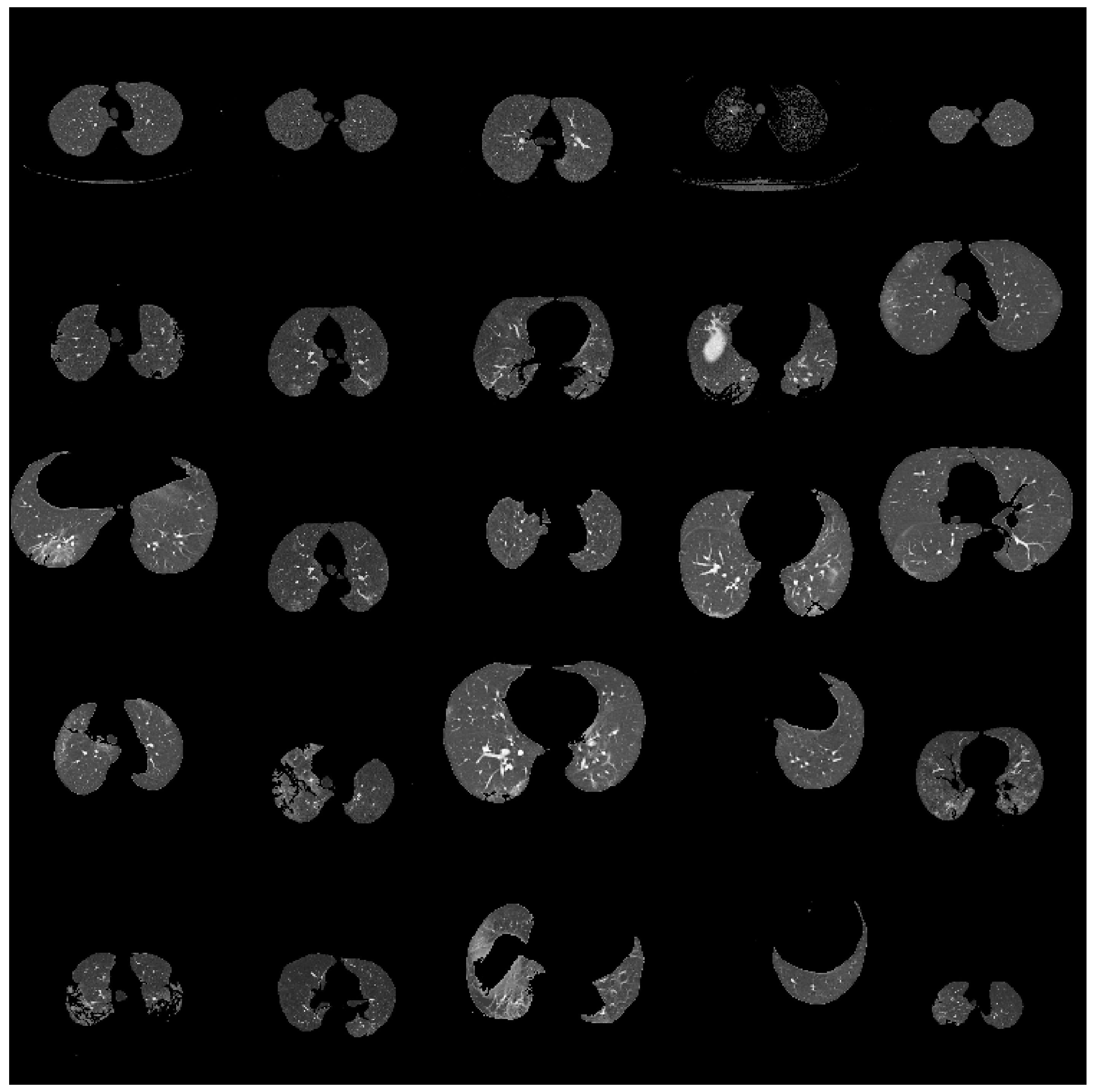}
            \caption{CT scan samples}
            \label{fig:SSL_F2232}
    \end{subfigure}
    \begin{subfigure}[b]{0.18\textwidth}
            \centering
            \includegraphics[width=\textwidth]{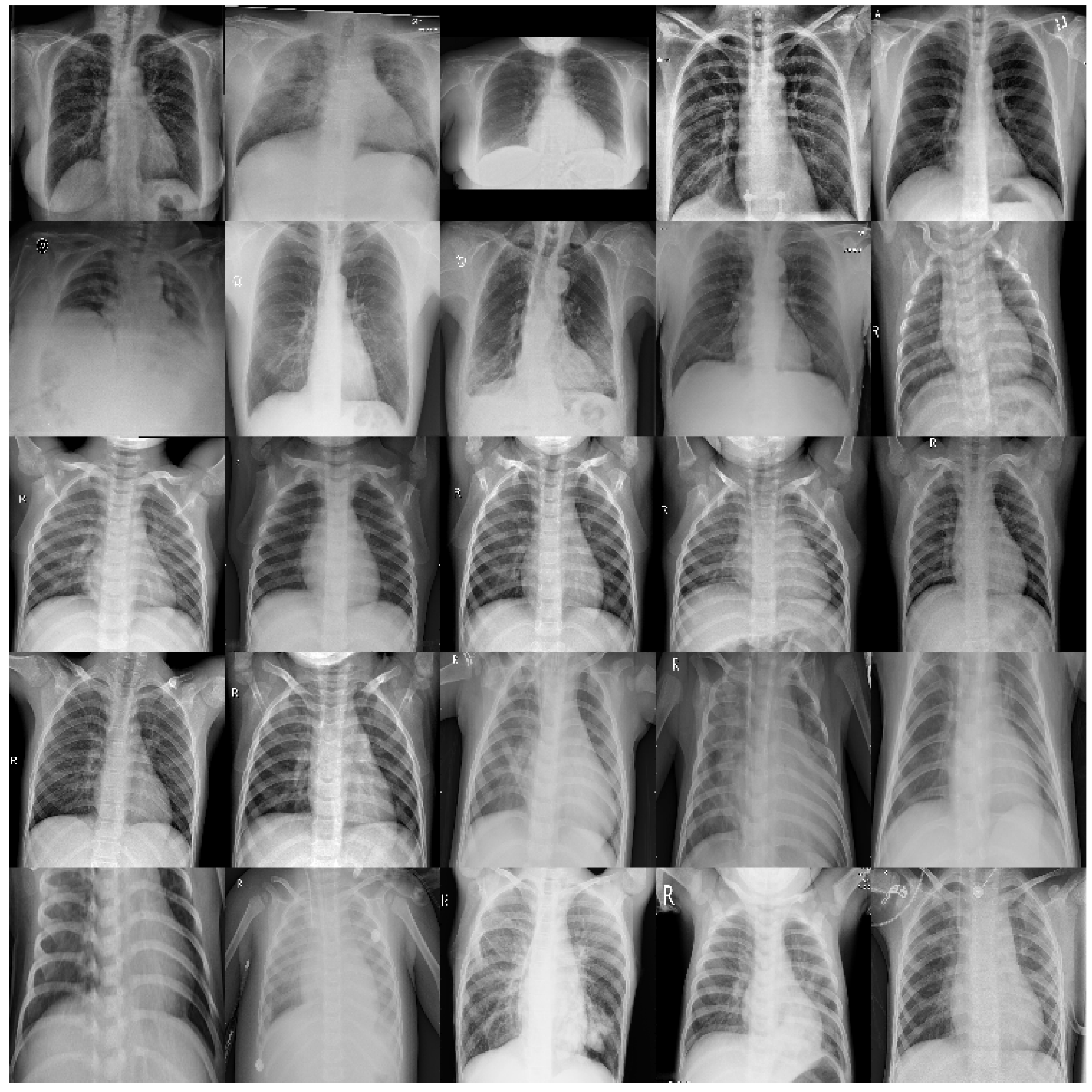}
            \caption{X-ray samples}
            \label{fig:SSL_F2324}
    \end{subfigure}  
    \caption{Some random image samples from the CT scan and X-ray datasets considered in our study.}\label{SmapleCO}
\end{figure}

In this study, two types of input image data were used: CT scan \cite{ning2020ictcf} \footnote{Sources:    \url{https://www.kaggle.com/azaemon/preprocessed-ct-scans-for-covid19}} and  and X-ray \footnote{Sources: \url{https://www.kaggle.com/prashant268/chest-xray-covid19-pneumonia}} images (see Table \ref{DATA}). Some random samples of the CT scan and X-ray datasets considered in this study are shown in Fig. \ref{SmapleCO}. The CT scan dataset has classes of data:  non-informative CT (NiCT), positive CT (pCT), and negative CT (nCT) images. The X-ray dataset also has three data classes: COVID-19, Normal, and Pneumonia images.

\begin{table}[]
\centering
\caption{Characteristics of the CT scan and X-ray datasets considered in our study.}
\begin{tabular}{lll}
\hline
Dataset & \# of Samples & \# of Classes \\ \hline
CT scan images &  19685 &  3 \\ \cline{2-3} 
X-ray images   &  6432  & 3 \\ \hline
\label{DATA}
\end{tabular}
\end{table}

\subsection{Experimental Results}
\label{Sec:sec:ER}
In this section, the experimental results are presented and discussed. Since we also considered the impact of UQ methods, our experiments have been conducted with and without applying them for detection of COVID-19 cases. In our first experiment, we compared five different machine learning models, including Random Forest (RF), Decision Trees (DT, max-depth=50, and n-estimators=200), Deep 1 (Simple CNN), Deep 2 (Multi-headed CNN), and our proposed model (feature fusion model).

\subsubsection{\textbf{COVID-19 classification without considering uncertainty}} 
\label{CLWOUQ}
First, we investigated the performance of the five considered classifiers (RF, DT, simple CNN, multi-headed CNN and our proposed feature fusion model) without considering uncertainty. The obtained results are presented in Tables and \ref{WIUQCT} and \ref{WIXR} for the CT scan and X-ray datasets, respectively. As shown in Table \ref{WIUQCT} our feature fusion model outperformed the other methods for the CT scan dataset, providing the accuracy of 99.136\%, and followed by simple CNN with the accuracy of 98.763\%. The obtained results also indicate that DT provided the weakest performance for the CT scan dataset among the five competing models. Fig. 1 in the Supplementary Material and Fig. \ref{CT11411} present the confusion matrices and the ROC curves obtained for the CT scan dataset without quantifying uncertainty, respectively.

\begin{table}[]
\scriptsize
\caption{Comparison of the results (given in \%) provided by different ML models for detecting COVID-19 cases for the CT scan dataset: Results without considering uncertainty.}
\begin{tabular}{lllll}
\hline
ML Model  & Precision & Recall & F-Measure & Accuracy \\ \hline
RF& 97.111 & 97.070 & 97.091 & 97.070  \\
DT  & 93.049 & 93.040 & 93.045 & 93.040 \\
Deep 1 (Simple CNN) &98.787 &98.763 &98.775 &98.763  \\
Deep 2 (Multi-headed CNN) & 98.599 & 98.577 & 98.588 & 98.577 \\
\textbf{Proposed (Fusion model)} & \textbf{99.137} & \textbf{99.136} & \textbf{99.136} & \textbf{99.136} \\ \hline
\label{WIUQCT}
\end{tabular}
\end{table}


\begin{figure*}[h!]
\centering
    \begin{subfigure}[b]{0.19\textwidth}
            \includegraphics[width=\textwidth]{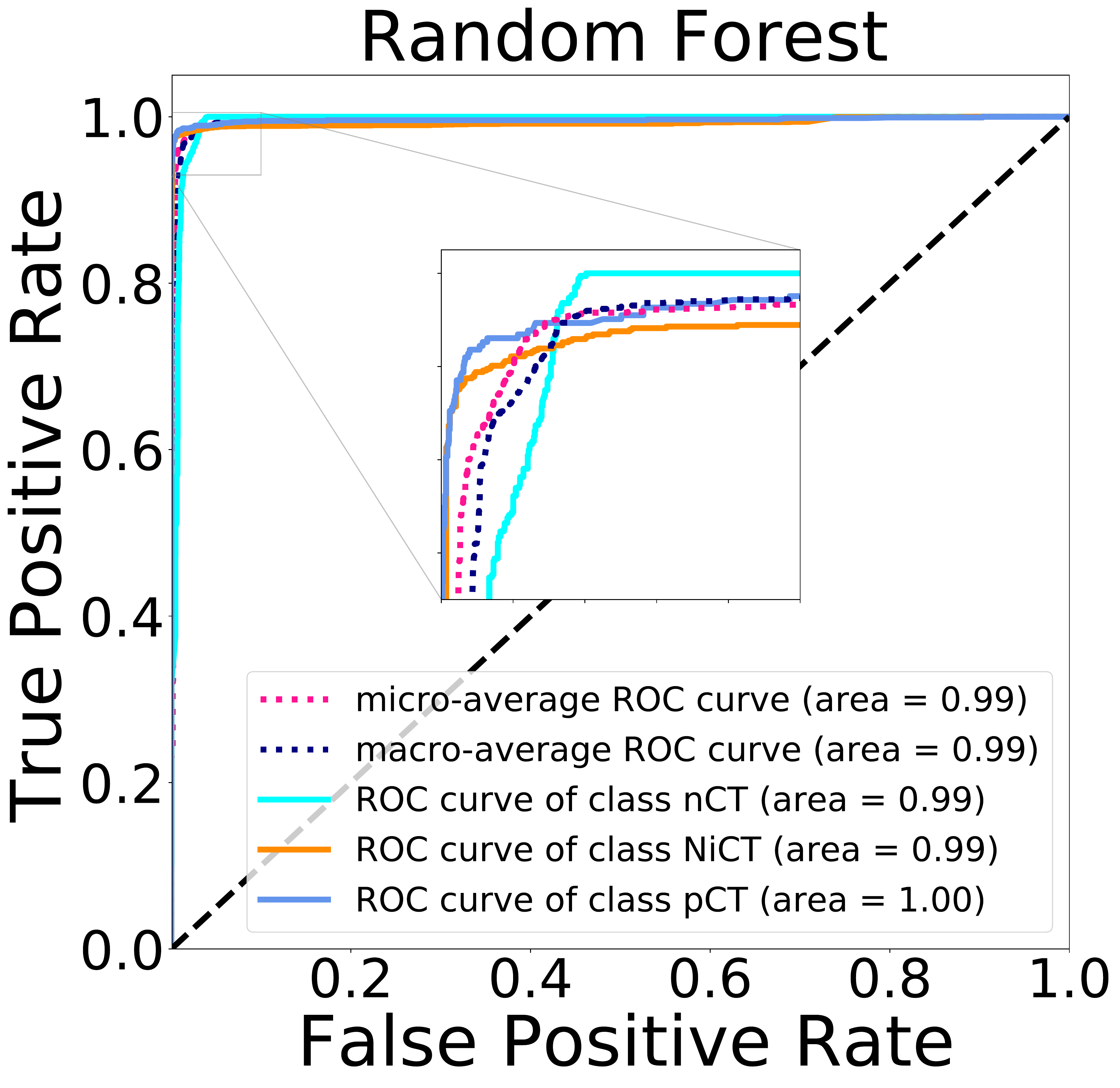}
            \caption{Random Forest}
            \label{fig:SSL_F11111}
    \end{subfigure}
    \begin{subfigure}[b]{0.19\textwidth}
            \centering
            \includegraphics[width=\textwidth]{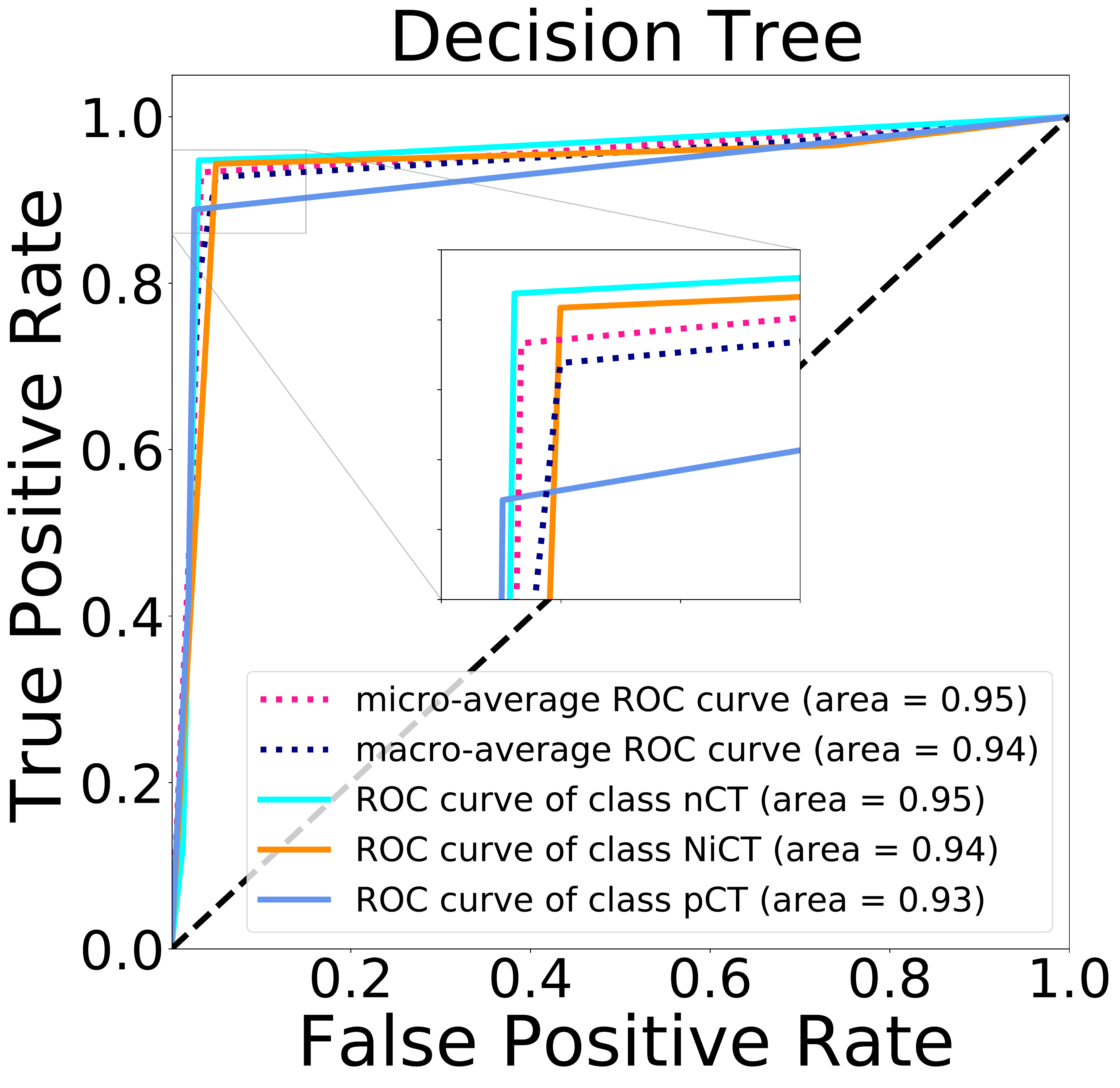}
            \caption{Decision Tree}
            \label{fig:SSL_F222}
    \end{subfigure} 
    \begin{subfigure}[b]{0.19\textwidth}
            \centering
            \includegraphics[width=\textwidth]{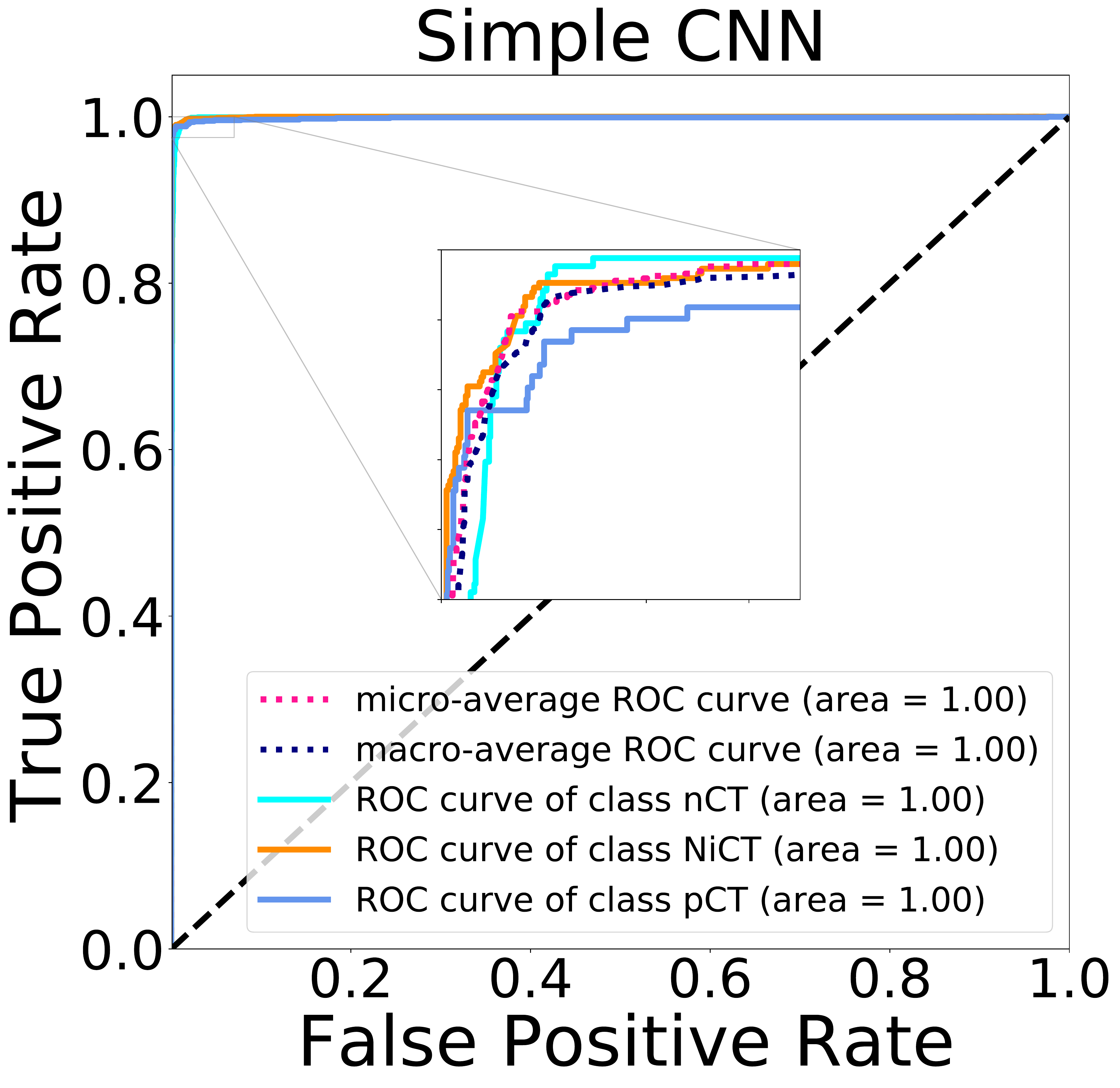}
            \caption{Simple CNN}
            \label{fig:SSL_F2232}
    \end{subfigure}
    \begin{subfigure}[b]{0.19\textwidth}
            \centering
            \includegraphics[width=\textwidth]{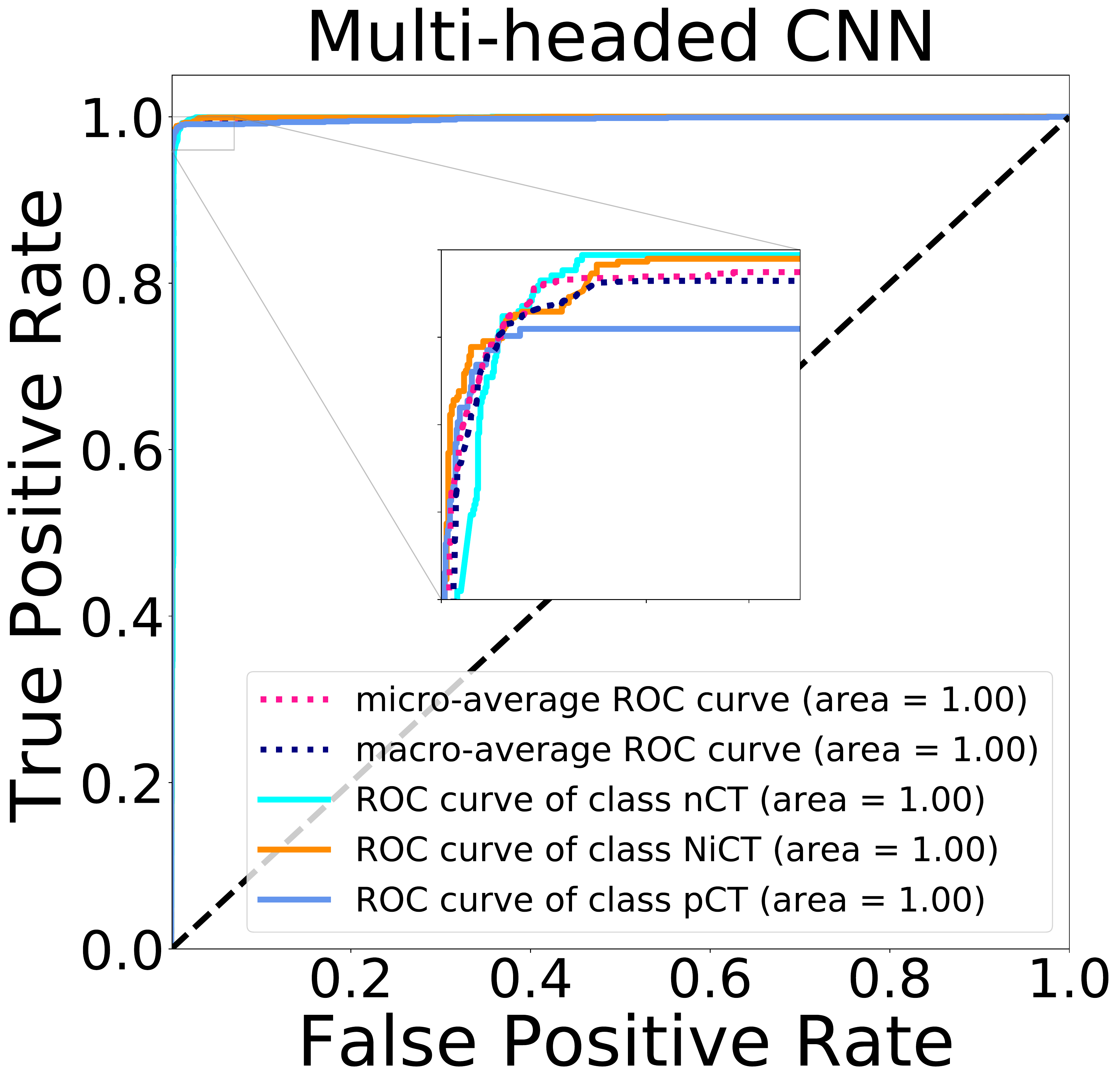}
            \caption{Multi-headed CNN}
            \label{fig:SSL_F2324}
    \end{subfigure}  
    \begin{subfigure}[b]{0.19\textwidth}
            \centering
            \includegraphics[width=\textwidth]{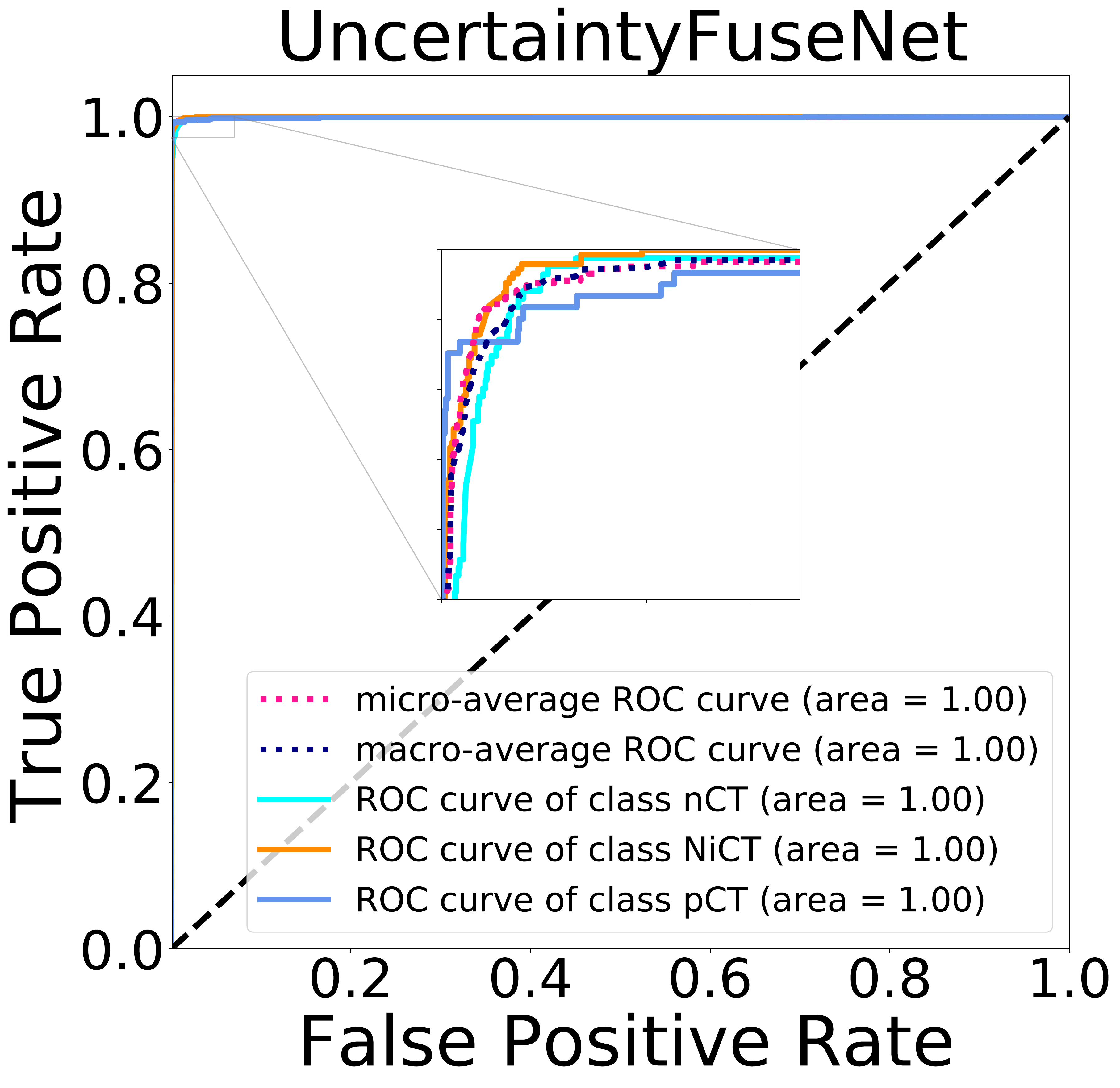}
            \caption{Fusion Model}
            \label{fig:SSL_F2324}
    \end{subfigure}
         \caption{ROC curves obtained for the five considered ML models for the CT scan data without quantifying uncertainty.}\label{CT11411}
\end{figure*}

To demonstrate the effectiveness of the proposed feature fusion model, the same five ML have been applied to analyze X-ray data. It can be observed from Table \ref{WIXR} that our feature fusion model performed much better than the other competing ML models, providing the accuracy of 97.127\%, followed by the multi-headed CNN model with the accuracy of 94.953\%. The traditional Decision Tree model provided much worse results for the X-ray data (the recall value of 84.006\%) than for the CT scan data (the recall value of 93.040\%). Fig. 2 in the Supplementary Material and Fig. \ref{RWIUQ} present the confusion matrices and the ROC curves obtained by the five ML models for the X-ray dataset without quantifying uncertainty. 
   
\begin{table}[h!]
\scriptsize
\caption{Comparison of the results (given in \%) provided by different ML models for detecting COVID-19 cases for the X-ray dataset: Results without considering uncertainty.}
\begin{tabular}{lllll}
\hline
ML Model & Precision & Recall & F-Measure & Accuracy \\ \hline
RF & 91.532 & 91.381 & 91.456 & 91.381 \\
DT  & 83.828 & 84.006 & 83.917 & 84.006 \\
Deep 1 (Simple CNN) & 93.847 & 93.167 & 93.506 & 93.167  \\
Deep 2 (Multi-headed CNN)  & 95.041 & 94.953 & 94.997 & 94.953  \\
DarkCovidNet & 95.752 & 95.729 & 95.741& 95.729  \\
\textbf{Proposed (Fusion model)} & \textbf{97.121} & \textbf{97.127} & \textbf{97.124} & \textbf{97.127} \\ \hline
\label{WIXR}
\end{tabular}
\end{table}


\begin{figure*}[h!]
\centering
    \begin{subfigure}[b]{0.19\textwidth}
            \includegraphics[width=\textwidth]{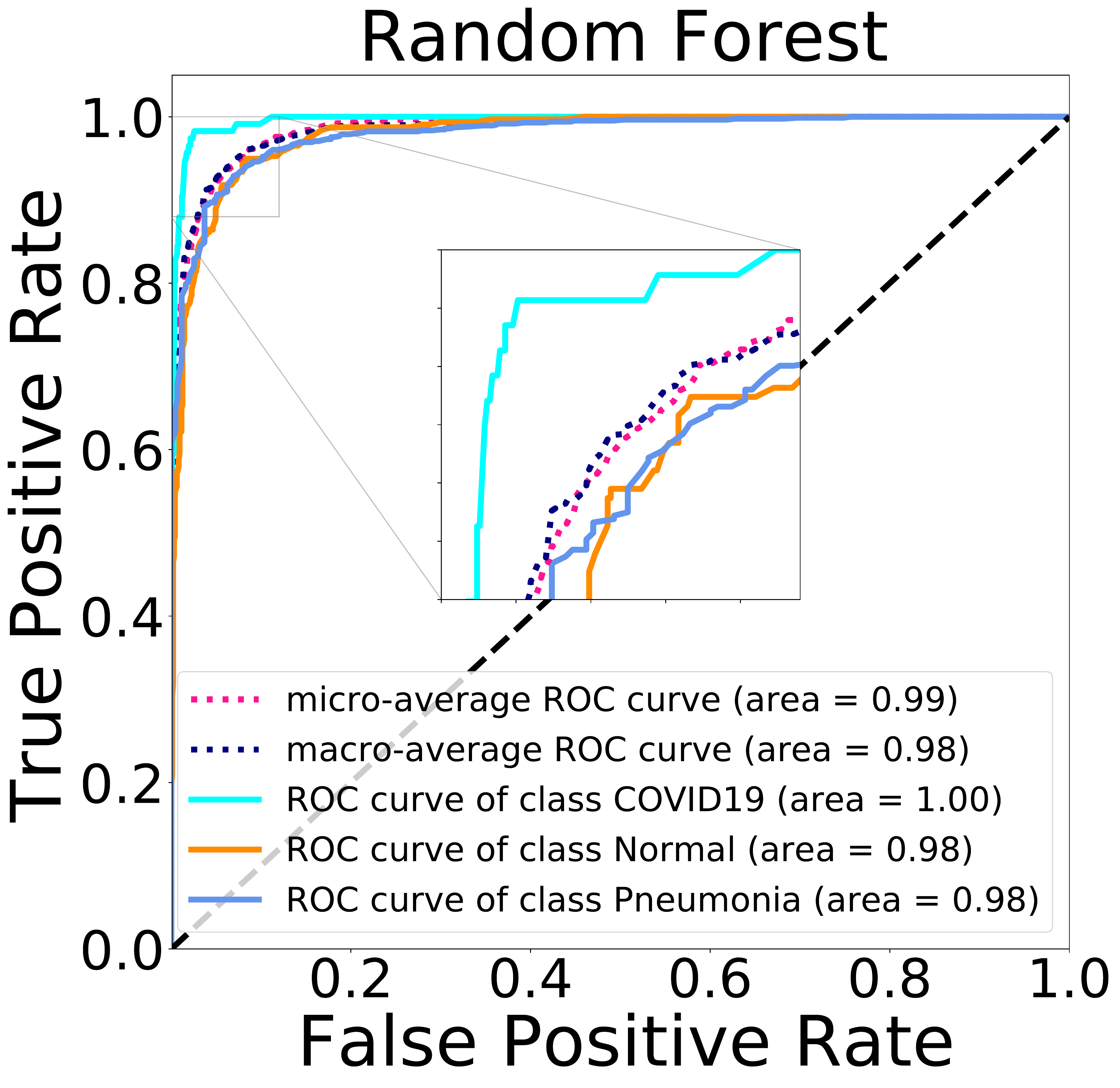}
            \caption{Random Forest}
            \label{fig:SSL_F11111}
    \end{subfigure}
    \begin{subfigure}[b]{0.19\textwidth}
            \centering
            \includegraphics[width=\textwidth]{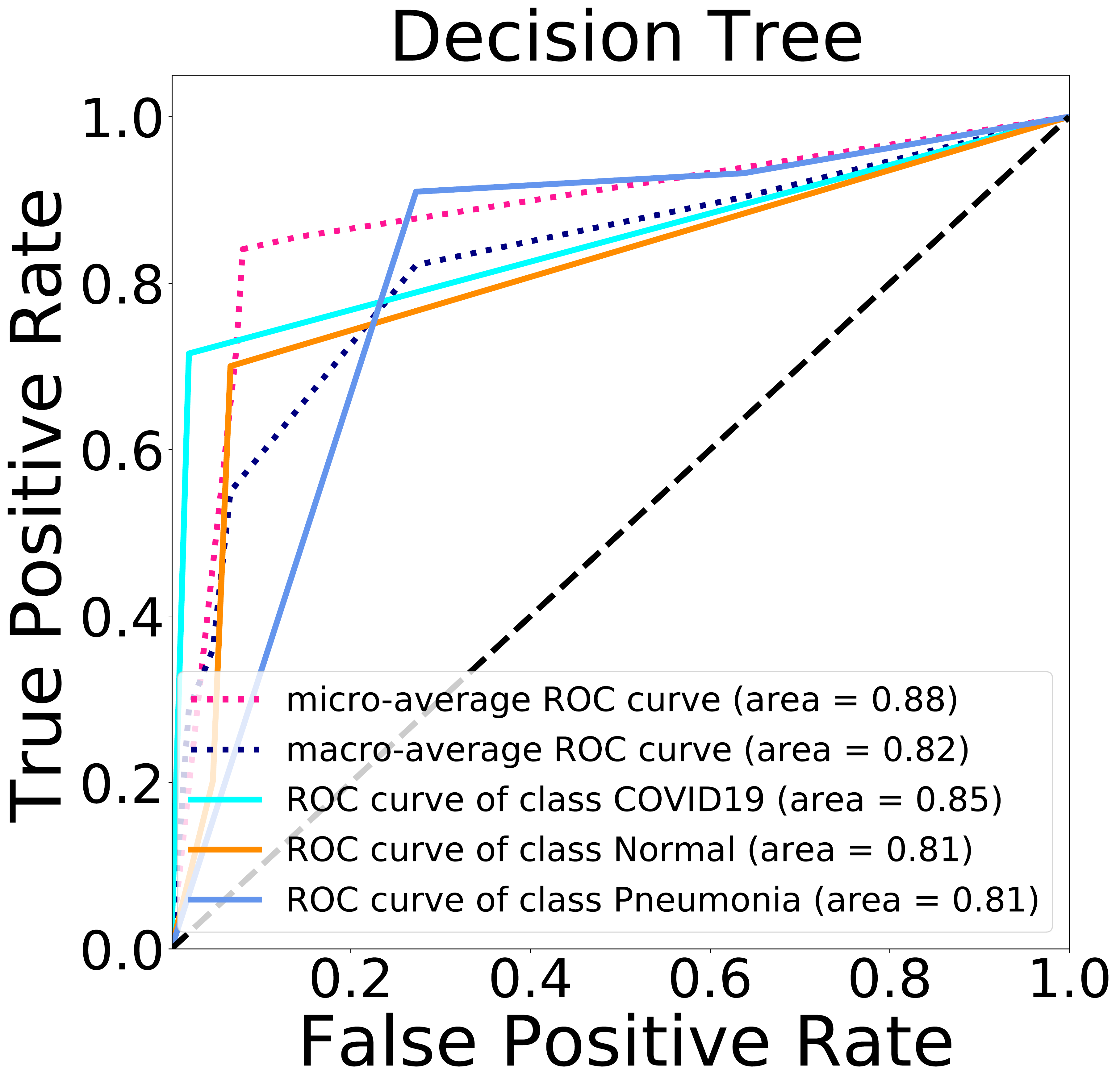}
            \caption{Decision Tree}
            \label{fig:SSL_F222120}
    \end{subfigure} 
    \begin{subfigure}[b]{0.19\textwidth}
            \centering
            \includegraphics[width=\textwidth]{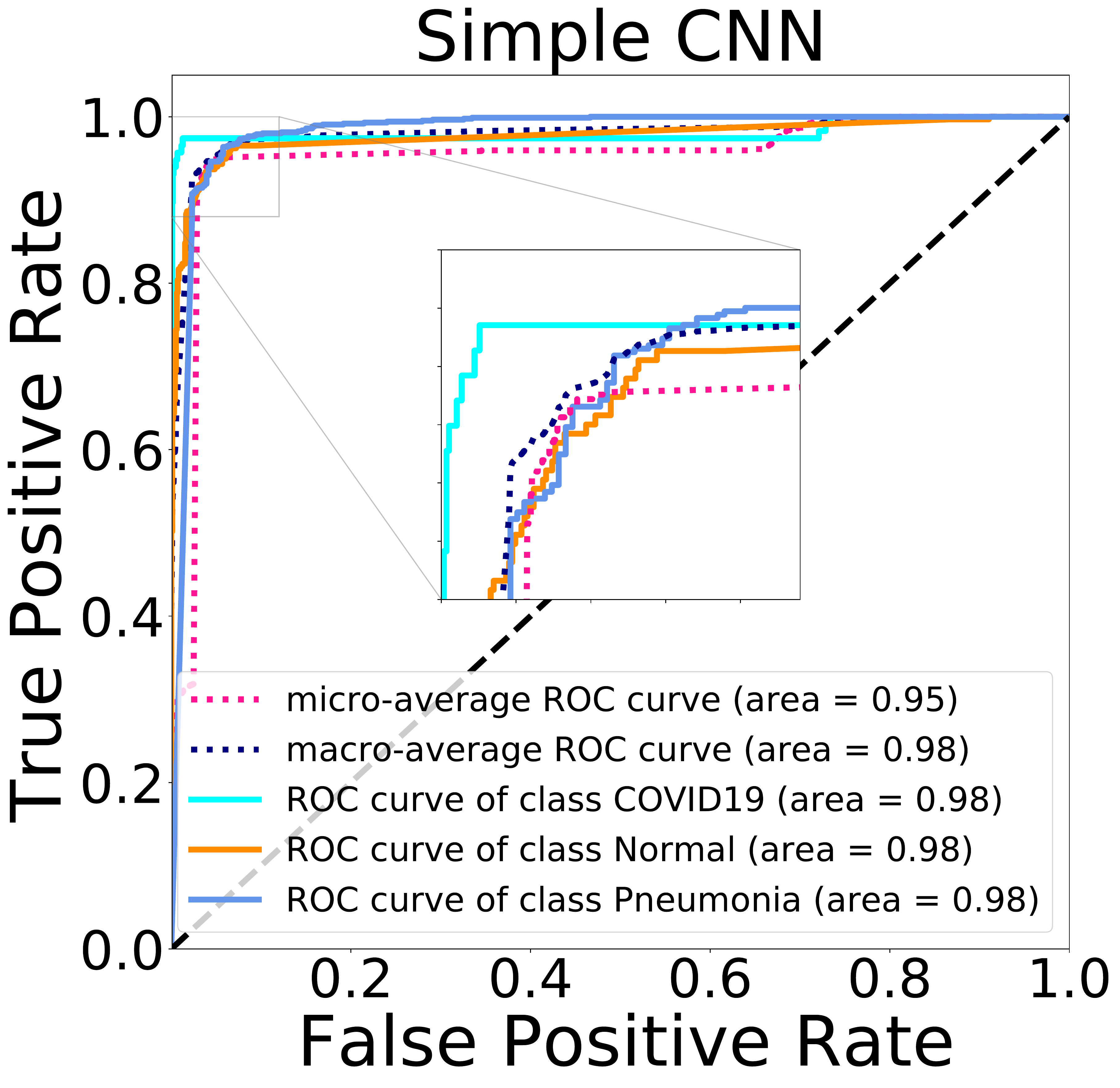}
            \caption{Simple CNN}
            \label{fig:SSL_F2232}
    \end{subfigure}
    \begin{subfigure}[b]{0.19\textwidth}
            \centering
            \includegraphics[width=\textwidth]{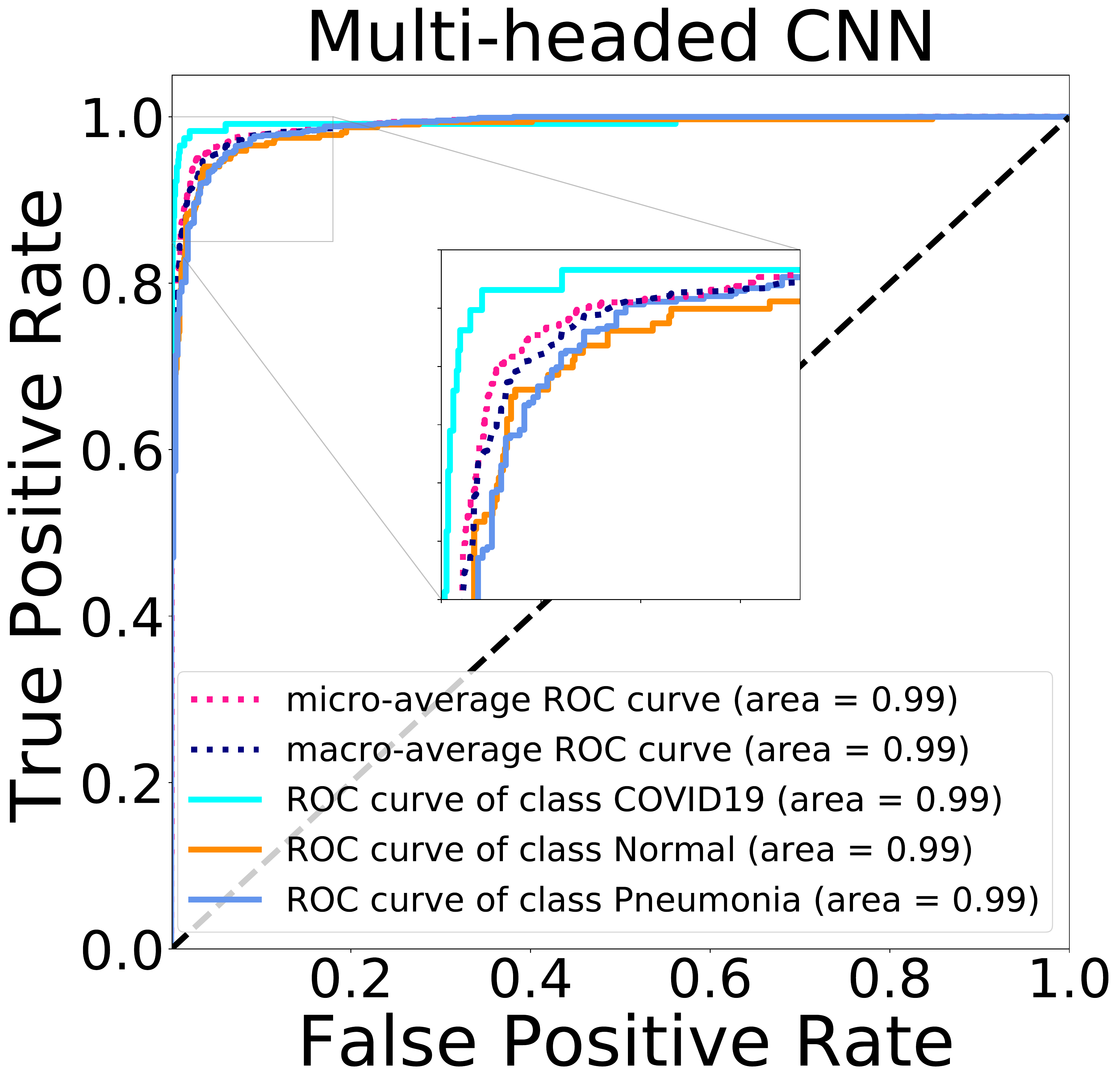}
            \caption{Multi-headed CNN}
            \label{fig:SSL_F2324}
    \end{subfigure}  
    \begin{subfigure}[b]{0.19\textwidth}
            \centering
            \includegraphics[width=\textwidth]{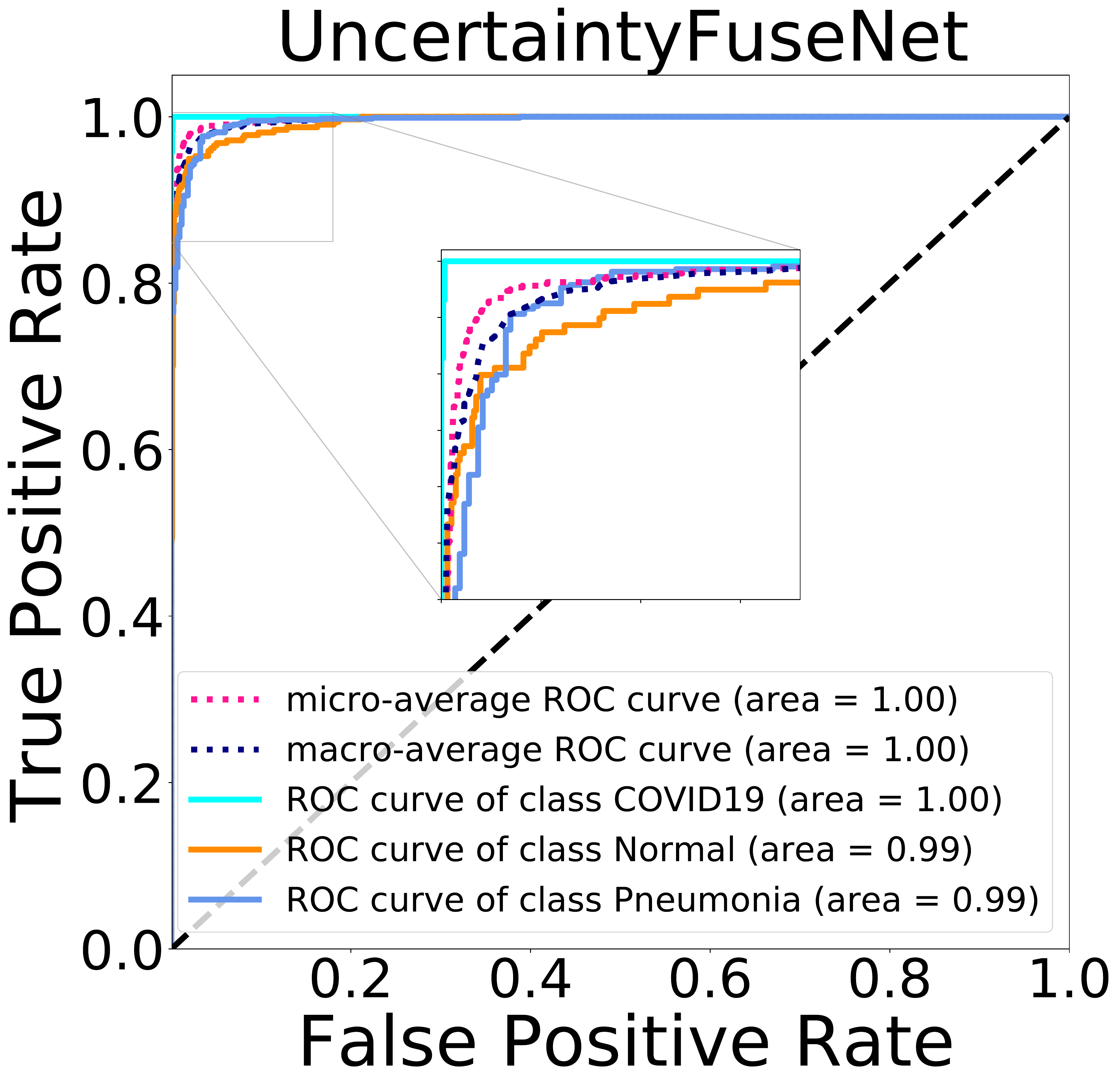}
            \caption{Fusion Model}
            \label{fig:SSL_F2324}
    \end{subfigure}
    \caption{ROC curves obtained for the five considered ML models for the X-ray data without quantifying uncertainty.}\label{RWIUQ}
\end{figure*}

\subsubsection{\textbf{COVID-19 classification considering uncertainty}}
\label{CLWUQ}
The results, discussed in the previous sub-section (\ref{CLWOUQ}), provided by our new feature fusion model are promising, suggesting that it can be used by clinical practitioners for automatic detection of COVID-19 cases. We believe that new efficient intelligent (\emph{i.e.} ML and DL) models to deal with COVID-19 data are urgently needed. At the same time, we firmly believe in the uncertainty estimates should accompany such intelligent models. To accomplish this, we applied the uncertainty quantification method, called EMC dropout, to estimate the uncertainty of our deep learning predictions. The EMC method was used in the framework of the Deep 1 (Simple CNN) and Deep 2 (Multi-headed CNN) models, and our proposed UncertaintyFuseNet model.

Table \ref{WUQCT} and Fig. 3 in the Supplementary Material (confusion matrices) and Fig. \ref{RWUQCT} (ROC curves) show the results provided by the three compared deep learning models considering uncertainty for the CT scan dataset. As shown in Table \ref{WUQCT}, our feature fusion model yielded a better classification performance compared to the Deep 1 and Deep 2 CNN-based models. UncertaintyFuseNet provided the accuracy value of 99.085\%, followed by the Deep 1 model with the accuracy value of 98.831\%, for the CT scan data. The results obtained using deep learning models with and without uncertainty quantification (UQ) reveal that our proposed feature fusion model with UQ method has had a slightly poorer performance than the model without UQ. The Deep 1 CNN model performed slightly better with UQ, while the Deep 2 CNN model performed slightly better without UQ.

\begin{table}[h!]
\scriptsize
\caption{Comparison of the results (given in \%) provided by the 3 DL models for detecting COVID-19 cases for the CT scan dataset: Results obtained with uncertainty quantification.}
\begin{tabular}{lllll}
\hline
DL Model & Precision & Recall & F-Measure & Accuracy \\ \hline
Deep 1 (Simple CNN) & 98.831 & 98.854& 98.843& 98.831  \\
Deep 2 (Multi-headed CNN) & 98.493 & 98.523 & 98.508 & 98.493 \\
\textbf{Proposed (Fusion model)} & \textbf{99.085} & \textbf{99.085} &  \textbf{99.085} & \textbf{99.085}  \\ \hline
\label{WUQCT}
\end{tabular}
\end{table}


\begin{figure*}[h!]
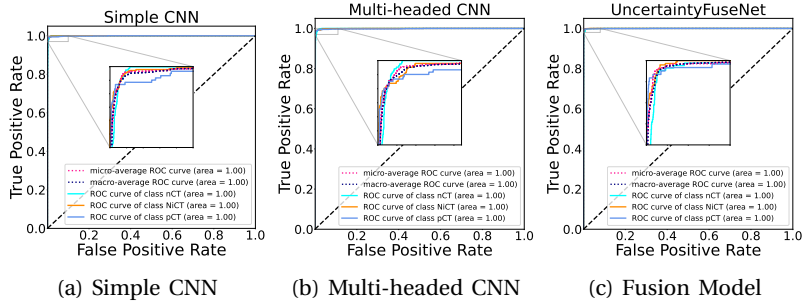

\centering
    \begin{subfigure}[b]{0.19\textwidth}
            \centering
            \includegraphics[width=\textwidth]{ROC-of-Simple-CNN-model-with-uncertainty-CT-scan.pdf}
            \caption{Simple CNN}
            \label{fig:SSL_F2232}
    \end{subfigure}
    \begin{subfigure}[b]{0.19\textwidth}
            \centering
            \includegraphics[width=\textwidth]{ROC-of-Multi-headed-model-with-uncertainty-CT-scan.pdf}
            \caption{Multi-headed CNN}
            \label{fig:SSL_F2324}
    \end{subfigure}  
    \begin{subfigure}[b]{0.19\textwidth}
            \centering
            \includegraphics[width=\textwidth]{ROC-of-Fusion-Model-with-Uncertainty-CT-scan.pdf}
            \caption{Fusion Model}
            \label{fig:SSL_F2324}
    \end{subfigure}
    \caption{ROC curves obtained for the three considered DL models for the CT scan data with uncertainty quantification.}\label{RWUQCT}
\end{figure*}

We also evaluated the performance of three considered DL models with uncertainty quantification on the X-ray dataset. The obtained statistics, confusion matrices, and the ROC curves for the three competing DL models applied are presented in Table \ref{WUQXR}, and Fig. 4 in the Supplementary Material (confusion matrices) and Fig. \ref{RWUQXR} (ROC curves), respectively. Our proposed feature fusion model achieved the best performance for COVID-19 detection using X-Ray dataset with an accuracy of 96.350\% compared to the simple CNN (accuracy of 95.263\%). For the X-ray data, the proposed UncertaintyFuseNet model outperformed the Deep 1 simple CNN model, but was slightly surpassed by the Deep 2 multi-headed CNN (see Table \ref{WUQXR}).
in the Supplementary Material (confusion matrices)

\begin{table}[h!]
\scriptsize
\caption{Comparison of the results (given in \%) provided by the 3 DL models for detecting COVID-19 cases for the X-ray dataset: Results obtained with uncertainty quantification.}
\begin{tabular}{lllll}
\hline
Method & Precison & Recall & F-Measure & Accuracy \\ \hline
Deep 1 (Simple CNN) & 95.263 & 95.354 & 95.309 & 95.263  \\
Deep 2 (Multi-headed) & 95.186 & 95.257 & 95.222 & 95.186   \\
DarkCovidNet & 97.460 & 97.4589 & 97.459& 97.458  \\
\textbf{Proposed (Fusion model)} & \textbf{96.350} & \textbf{96.370} & \textbf{96.360} & \textbf{96.350} \\ \hline
\label{WUQXR}
\end{tabular}
\end{table}

\begin{figure*}[h!]
\centering
    \begin{subfigure}[b]{0.19\textwidth}
            \centering
            \includegraphics[width=\textwidth]{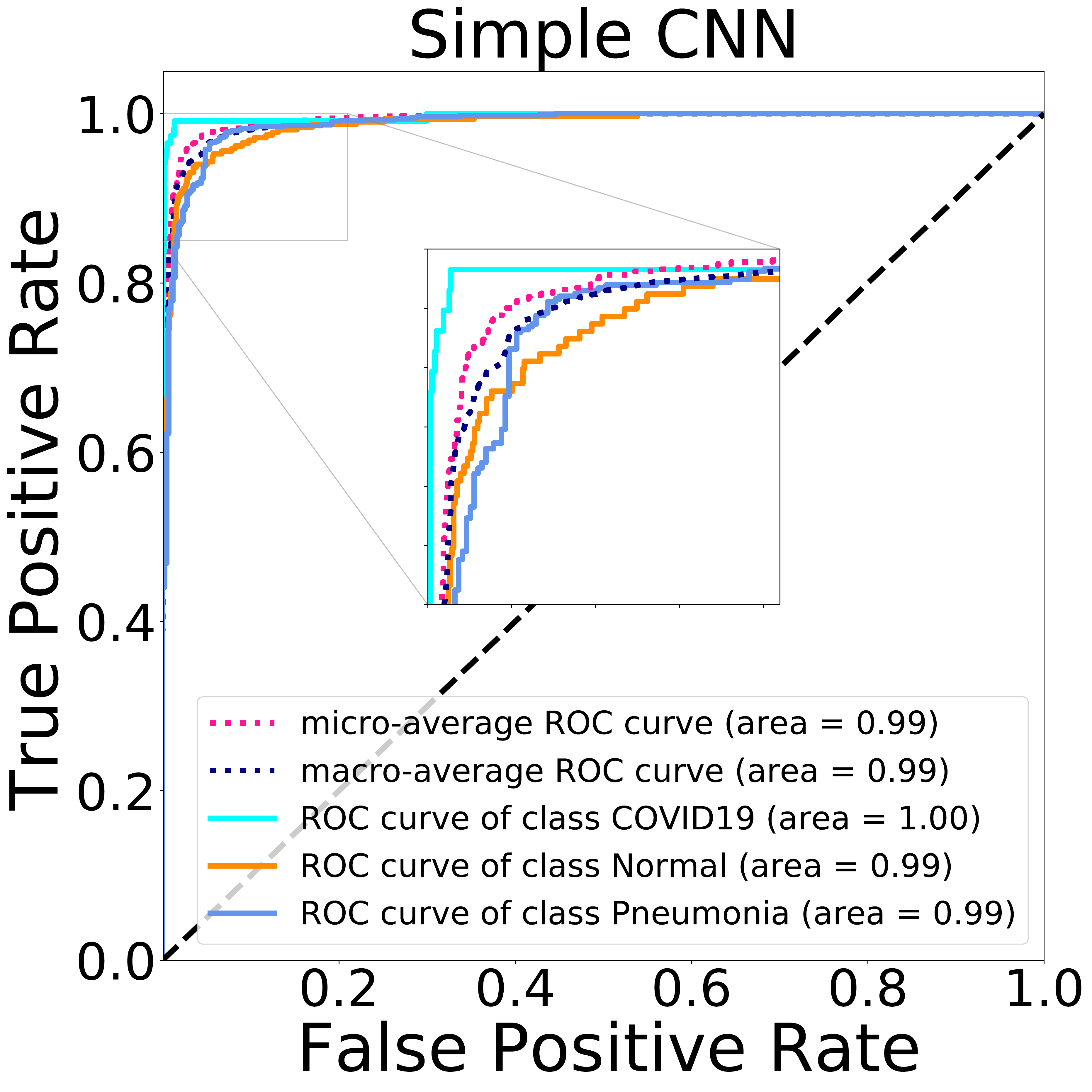}
            \caption{Simple CNN}
            \label{fig:SSL_F2232}
    \end{subfigure}
    \begin{subfigure}[b]{0.19\textwidth}
            \centering
            \includegraphics[width=\textwidth]{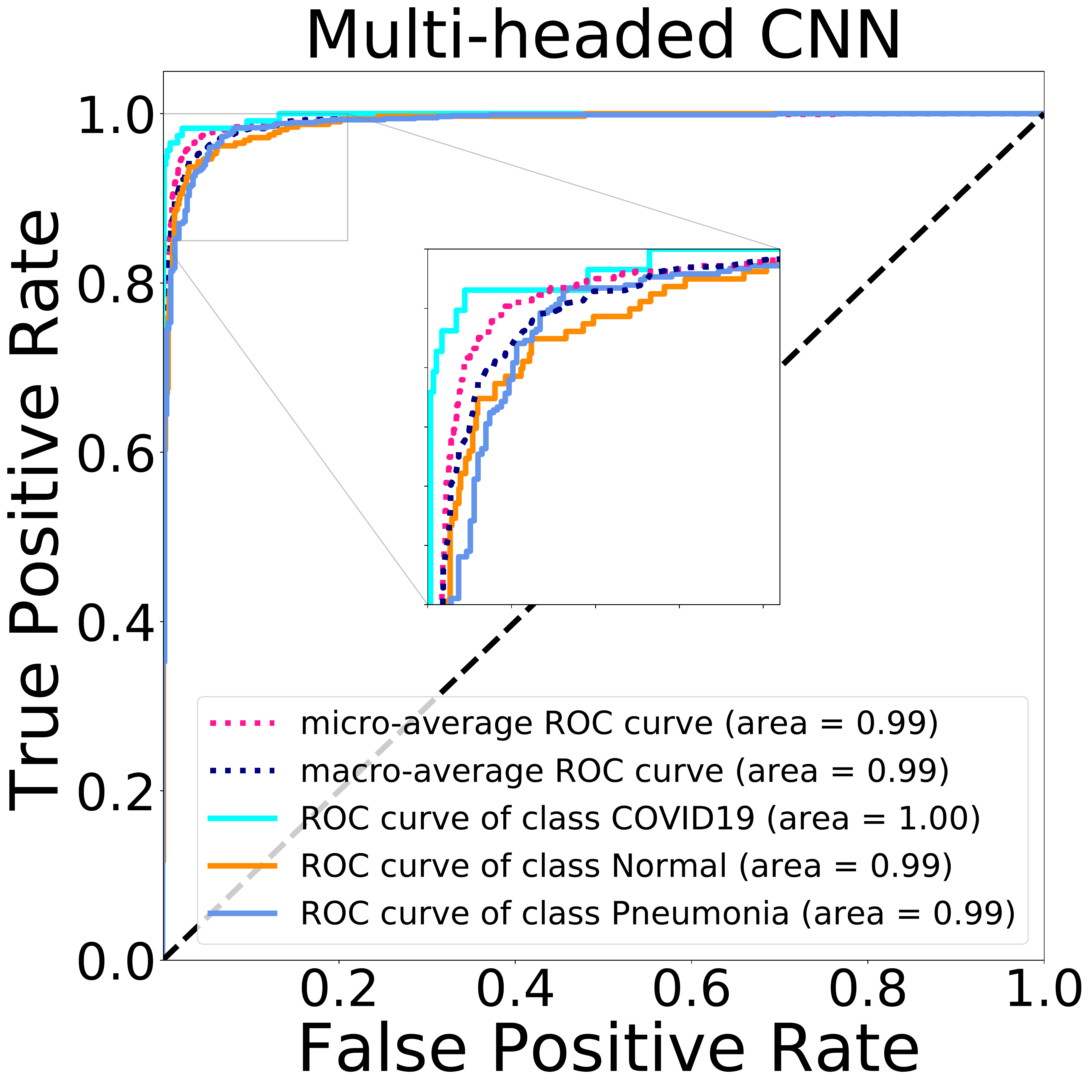}
             \caption{Multi-headed CNN}
            \label{fig:SSL_F2324}
    \end{subfigure}  
    \begin{subfigure}[b]{0.19\textwidth}
            \centering
            \includegraphics[width=\textwidth]{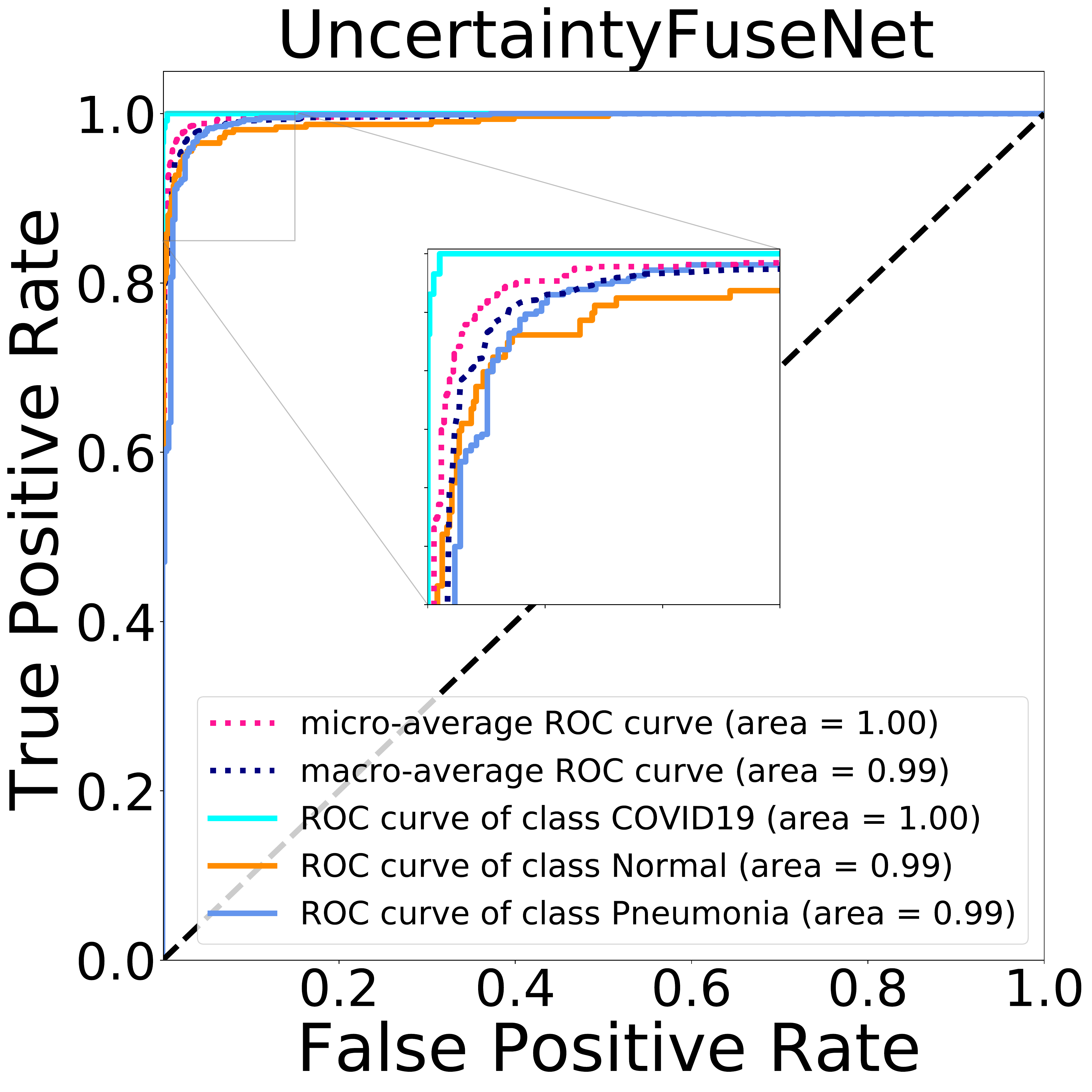}
            \caption{Fusion Model}
            \label{fig:SSL_F2324}
    \end{subfigure}
    \caption{ROC curves obtained for the three considered DL models for the X-ray data with uncertainty quantification.}\label{RWUQXR}
    \end{figure*}
    
\subsection{Robustness Against Noise}
\label{Sec:sec:RAN}
An individual visual system is significantly robust against a wide variety of natural noises and corruptions occurring in the nature such as snow, fog or rain~\cite{rusak2020simple}. However, the overall performance of various modern image and speech recognition systems is greatly degraded when evaluated using previously unseen noises and corruptions. Thus, conducting robustness tests for considered ML and DL models can be necessary to reveal their level of stability against noise. In this  study, the robustness of the applied deep learning models against noise has been investigated. \\
We added different noise variables to both CT scan and X-ray datasets to evaluate the performance of Simple CNN, Multi-headed CNN and our proposed feature fusion model. Gaussian noise variables with different standard deviations (STD) were generated. The generated STD values were the following: 0.0001, 0.001, 0.01, 0.1, 0.2, 0.3, 0.4, 0.5, and 0.6, whereas the value of Mean was equal to 0. Our simulation results obtained for the CT scan and X-ray datasets are presented in Tables \ref{NOICT} and \ref{NOIXR}, respectively. It may be noted from Table \ref{NOICT} (CT scan data results) that both Simple CNN and Multi-headed CNN models did not perform well with noisy data compared to our feature fusion model. The results reported in Table \ref{NOICT} indicate that the values of all metrics computed for Simple and Multi-headed CNNs decrease dramatically as the level of noise increases. In contrast, our feature fusion model has been much more robust against noise according to all metrics considered.

\begin{table}[h!]
\centering
\scriptsize
\caption{Robustness against noise results (given in \%) provided by the 3 compared DL models for detecting COVID-19 cases for the CT scan dataset. Here, $Noise$ $STD = \frac{\sigma_{\epsilon}}{\sigma_{0}}$, where $\sigma_{\epsilon}$ is the mean of the noise and $\sigma_{0}$ is standard deviation of the noise.}
\label{NOICT}
\begin{tabular}{llllll}
\hline
DL Model & Noise STD & Precision & Recall & F-Measure & Accuracy \\ \hline
\multirow{9}{*}{\shortstack{Deep 1\\ (Simple CNN)}} 
                                        & 0.0001  & 98.852 & 98.831 & 98.842 & 98.831  \\
                                        & 0.001& 98.868 & 98.848 & 98.858 & 98.848  \\
                                        & 0.01& 98.754 & 98.730 & 98.742 & 98.730 \\
                                        & 0.1& 95.785 & 95.614 & 95.700 & 95.614 \\
                                        & 0.2  &  89.270 & 87.284 & 88.266 & 87.284  \\
                                        & 0.3  &  86.370 & 82.593 & 84.439 & 82.593  \\
                                        & 0.4  &  82.628 & 75.194 & 78.736 & 75.194  \\
                                        & 0.5  &  78.303 & 64.053 & 70.465 & 64.053  \\
                                        & 0.6 & 75.770 & 57.534 & 65.405 & 57.534 \\ \hline
\multirow{9}{*}{\shortstack{Deep 2\\ (Multi-headed)}} 
                                        & 0.0001  & 98.526 & 98.493 & 98.509 & 98.493  \\
                                        &  0.001 &  98.524 & 98.493 & 98.508 & 98.493 \\
                                        &0.01   & 98.558 & 98.526 & 98.542 & 98.526  \\
                                        & 0.1 &  93.447 & 92.871 & 93.158 & 92.871 \\
                                        & 0.2  &  86.943 & 83.423 & 85.147 & 83.423  \\
                                        & 0.3  & 80.168 & 69.065 & 74.203 & 69.065 \\
                                        & 0.4  & 76.032 & 58.465 & 66.102 & 58.465  \\
                                        & 0.5  &  73.837 & 54.690 & 62.837 & 54.690 \\
                                        & 0.6 & 72.914 & 53.149 & 61.482 & 53.149 \\ \hline
\multirow{9}{*}{\shortstack{\textbf{Proposed} \\\textbf{(Fusion model)}}} 
                                        &  0.0001 &  99.085 & 99.085 & 99.085 & 99.085 \\
                                        &  0.001 &  99.119 & 99.119 & 99.119 & 99.119 \\
                                        &  0.01 & 99.194 & 99.187 & 99.190 & 99.187 \\
                                        & 0.1 &  99.098 & 99.085 & 99.092 & 99.085 \\
                                        & 0.2 &  98.828 & 98.814 & 98.821 & 98.814\\
                                        & 0.3  &  98.109 & 98.086 & 98.097 & 98.086 \\
                                        & 0.4  &  96.956 & 96.884 & 96.920 & 96.884 \\
                                        & 0.5  &  96.201 & 96.088 & 96.145 & 96.088 \\
                                        & 0.6  &  95.804 & 95.665 & 95.734 & 95.665  \\ \hline
\end{tabular}
\end{table}

Table \ref{NOIXR} reports the performance of the three selected deep learning models under different noise conditions for the X-ray dataset. Both Simple CNN and Multi-headed CNN did not perform well in this context, whereas our new model was usually much more robust against noise. It should be noted that our feature fusion model performed better for the CT scan data than for the X-ray data.

\begin{table}[h!]
\centering
\scriptsize
\caption{Robustness against noise results (given in \%) provided by the 3 compared DL models for detecting COVID-19 cases for the X-ray dataset.}
\label{NOIXR}
\begin{tabular}{llllll}
\hline
DL Model & Noise STD & Precision & Recall & F-Measure & Accuracy \\ \hline
\multirow{9}{*}{\shortstack{ Deep 1\\ (Simple CNN)}}
                                        & 0.0001 & 95.408 & 95.341 & 95.375 & 95.341  \\
                                        &  0.001 &  95.408 & 95.341 & 95.341 & 95.341 \\
                                        & 0.01  & 95.338 & 95.263 & 95.301 & 95.263 \\
                                        &  0.1 &  94.554 & 94.254 & 94.404 & 94.254 \\
                                       & 0.2  &  91.540 & 89.285 & 90.398 & 89.285 \\
                                        & 0.3  &  88.534 & 82.065 & 85.176 & 82.065 \\
                                        &   0.4&  85.770 & 73.136 & 78.951 & 73.136 \\
                                        & 0.5  &  84.294 & 64.518 & 73.092 & 64.518 \\
                                        &  0.6 &  82.545 & 57.375 & 67.696 & 57.375\\ \hline

\multirow{9}{*}{\shortstack{Deep 2\\ (Multi-headed)}} 
                                        & 0.0001 & 95.474 & 95.419 & 95.446 & 95.419   \\
                                        &  0.001 &  95.188 & 95.108 & 95.148 & 95.108 \\
                                        &  0.01 & 95.404 & 95.341 & 95.372 & 95.341  \\
                                        &  0.1 &  93.922 & 93.322 & 93.621 & 93.322 \\
                                       &  0.2&  88.861 & 82.453 & 85.537 & 82.453 \\
                                        & 0.3  &  83.781 & 58.074 & 68.598 & 58.074\\
                                        &  0.4&  82.207 & 40.062 & 53.871 & 40.062 \\
                                        & 0.5  &  81.750 & 31.521 & 45.499 & 31.521 \\
                                        & 0.6  &  81.366 & 27.639 & 41.262 & 27.639 \\ \hline
\multirow{9}{*}{\shortstack{\textbf{Proposed }\\\textbf{(Fusion model)}}}
                                        & 0.0001 &  96.498 & 96.506 & 96.502 & 96.506  \\
                                        &  0.001 &  96.568 & 96.583 & 96.576 & 96.583 \\
                                        & 0.01  & 96.492 & 96.506 & 96.499 & 96.506  \\
                                        &  0.1 & 96.363 & 96.350 & 96.357 & 96.350 \\
                                       & 0.2  &  94.403 & 94.254 & 94.329 & 94.254  \\
                                        & 0.3  & 91.769 & 91.071 & 91.418 & 91.071 \\
                                        &   0.4&  88.225 & 85.714 & 86.951 & 85.714 \\
                                        & 0.5  &  84.181 & 78.804 & 81.404 & 78.804\\
                                        &  0.6 & 81.082 & 68.322 & 74.157 & 68.322\\ \hline
\end{tabular}
\end{table}
This stage of the experiments was necessary to demonstrate the stability of the applied models against noise. Our results clearly indicate that the proposed feature fusion model is robust against noise for both considered types of image data: CT scan and X-ray images.
\subsection{Unknown Data Detection}
\label{Sec:sec:UNKN}
\begin{figure}[h]
\centering
            \includegraphics[width=0.075\textwidth]{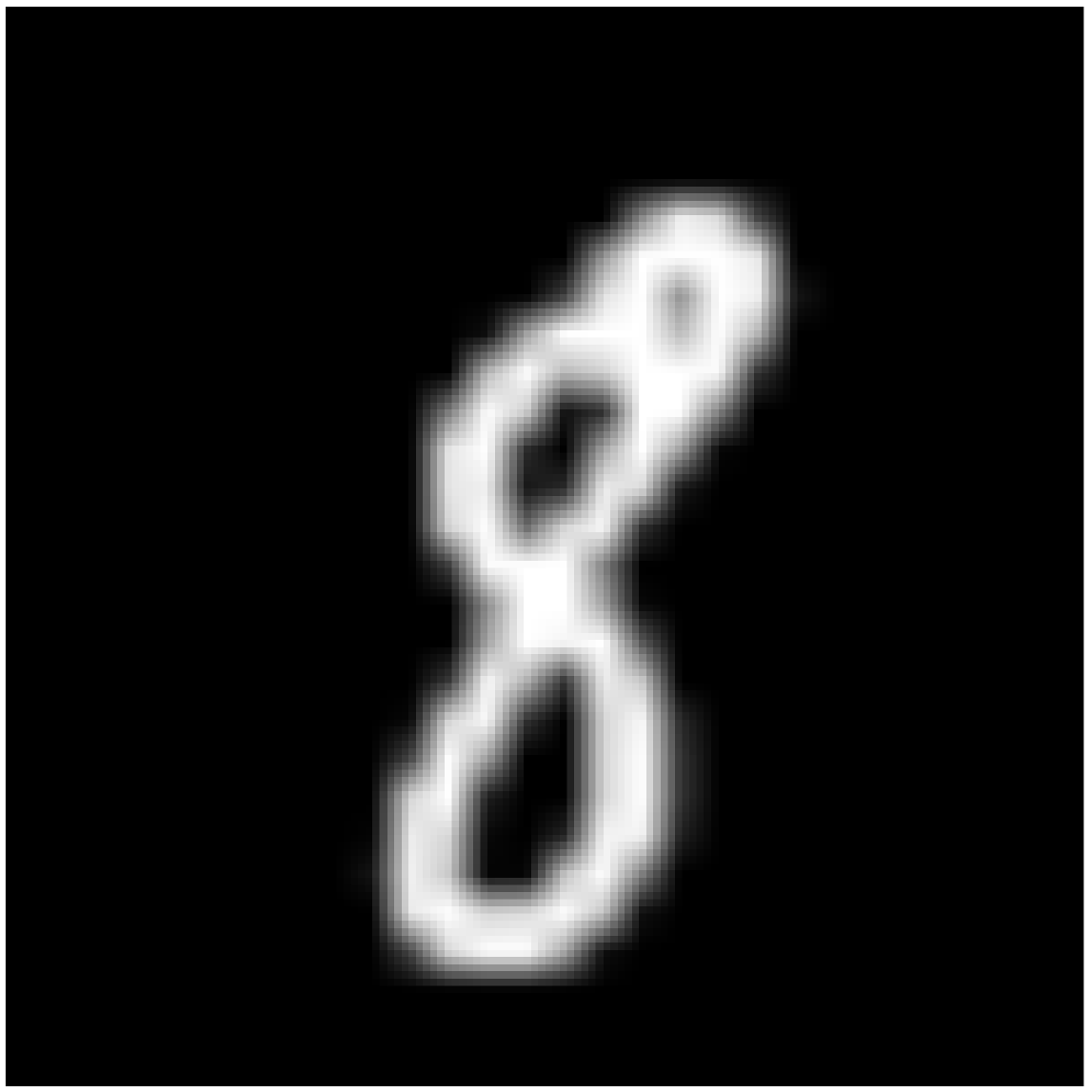}
    \caption{The MNIST sample image fed to the deep learning models as an unknown sample.}\label{MNIST}
\end{figure}

In this sub-section, we evaluate the performance of deep learning models when they are fed by unknown images. In these experimental settings the models either do not know or cannot clearly estimate the uncertainty of its predictions. To perform this evaluation, we fed the DL models being compared with one sample image from the well-known MNIST dataset (see Fig. \ref{MNIST}). The mean and the STD values of the Simple CNN model, Multi-headed CNN model and our proposed feature fusion model are reported in Table \ref{UNKN}. The obtained results indicate that our feature fusion model showed its uncertainty towards unknown data much better than the two other DL models.

We fed the MNIST sample image presented in Fig. \ref{MNIST} to the three deep learning models trained on CT scan and X-ray datasets, and then predicted the class of this unknown image sample. 

\begin{table*}[]
\centering
\caption{Unknown image class detection by Simple CNN, Multi-headed CNN and our proposed feature fusion model when fed with the image presented in Fig. \ref{MNIST}.} \label{UNKN}
\begin{tabular}{llllllll}
\hline
\multicolumn{2}{c}{\multirow{2}{*}{DL Model}}                                            & \multicolumn{3}{c}{CT scan} & \multicolumn{3}{c}{X-ray}     \\ \cline{3-8} 
\multicolumn{2}{c}{}                                            
& nCT     & NiCT     & pCT    & COVID-19 & Normal & Pneumonia \\ \hline
\multirow{2}{*}{\begin{tabular}[c]{@{}l@{}}Deep 1 \\ (Simple CNN)\end{tabular}}   
& Mean &0.02 & 0.98  &0.0  &0.57 &0.15 &  0.28\\ \cline{2-8} 
& STD  &0.10 &0.10  & 0.01 &0.39 &0.26 &0.35 \\ \hline
\multirow{2}{*}{\begin{tabular}[c]{@{}l@{}}Deep 2 \\(Multi-headed CNN)\end{tabular}} 
& Mean &0.05 &0.30  &0.65  &0.68  & 0.22  &0.10  \\ \cline{2-8} 
& STD  &0.19 & 0.43 &0.45 &0.32 & 0.27 & 0.16 \\ \hline
\multirow{2}{*}{\begin{tabular}[c]{@{}l@{}} \textbf{Proposed}\\ \textbf{Fusion Model}\end{tabular}}  
& Mean &\textbf{0.56} &\textbf{0.0}  & \textbf{0.44} & \textbf{0.41} &\textbf{0.59}  & \textbf{0.0}  \\ \cline{2-8} 
& STD  & \textbf{0.50}& \textbf{0.07} &\textbf{0.50}  & \textbf{0.49} & \textbf{0.49} & \textbf{0.0}  \\ \hline
\end{tabular}
\end{table*}

Estimating uncertainty of traditional machine learning and deep learning models using different UQ methods is vital during critical predictions such as medical case studies. Ideally, the applied ML models should be able to capture a portion of both epistemic and aleatoric uncertainties. In this study, we applied a new feature fusion model to classify two types of medical data: CT scan and X-ray images. Table \ref{UNKN} reports the \textit{Mean} and the \textit{STD} values of the three considered deep learning models applied to unknown data. It should be noted that the \textit{Mean} value accounts for the model's prediction and \textit{STD} accounts for its uncertainty. As reported in Table \ref{UNKN}, our model usually provides zero (or close to zero) values of \textit{Mean} and \textit{STD} for one of the image classes (for both CT scan and X-ray image types).


\section{Discussion}
\label{Sec:DI}
\begin{figure*}[h!]
\centering
    \begin{subfigure}[b]{0.23\textwidth}
            \centering
            \includegraphics[width=\textwidth]{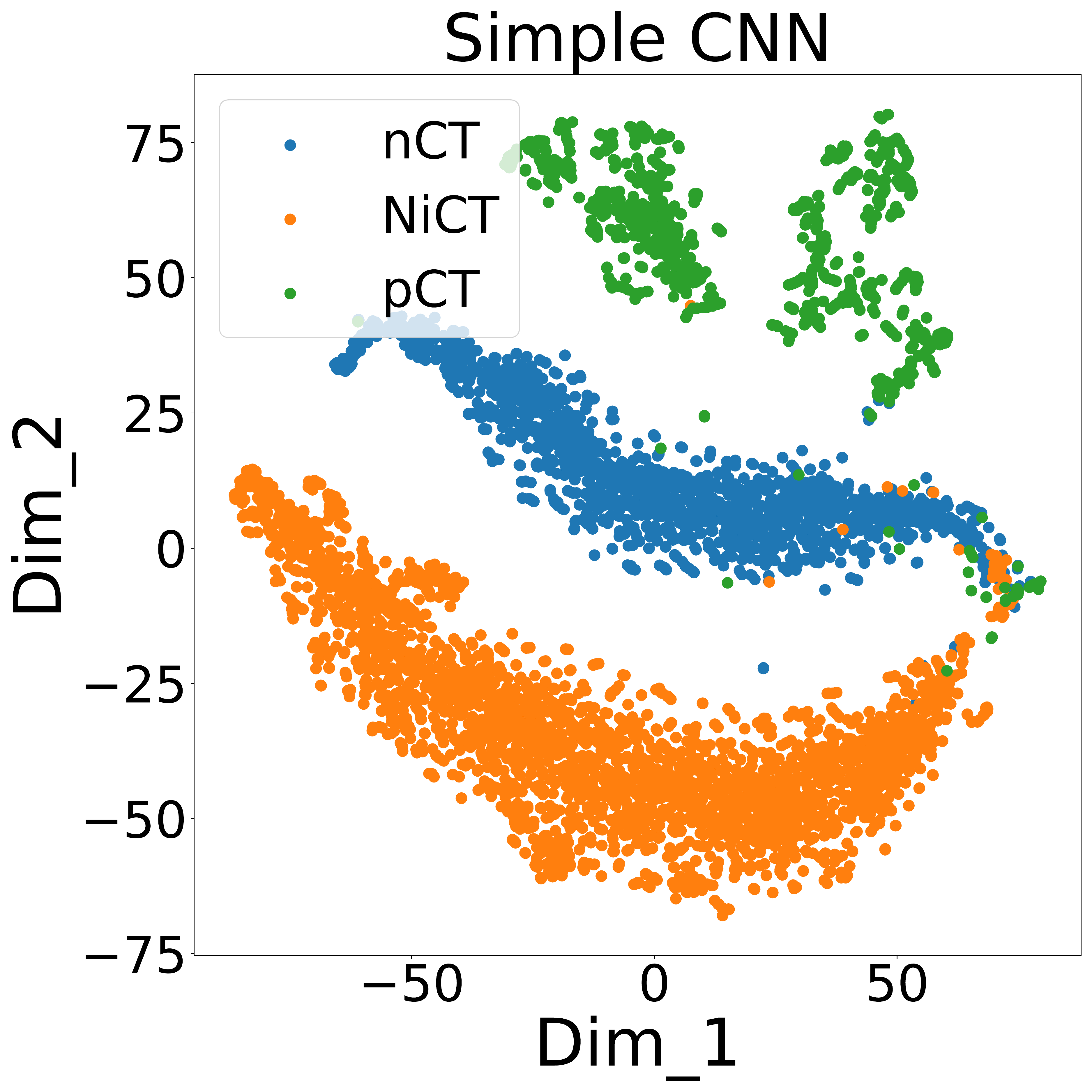}
            \caption{Simple CNN without UQ}
            \label{fig:SSL_F2232}
    \end{subfigure}
    \begin{subfigure}[b]{0.23\textwidth}
            \centering
            \includegraphics[width=\textwidth]{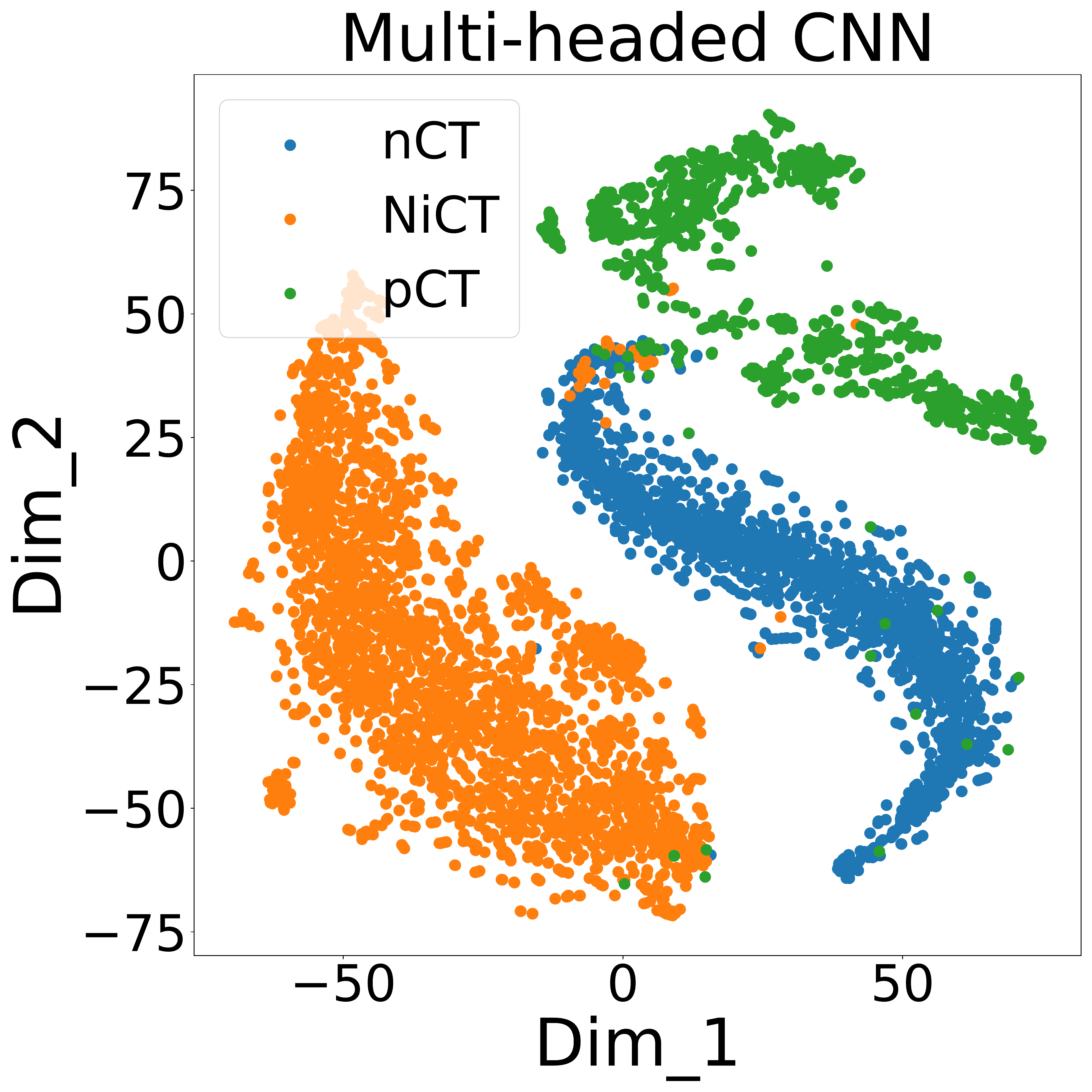}
            \caption{Multi-headed without UQ}
            \label{fig:SSL_F2324}
    \end{subfigure}  
    \begin{subfigure}[b]{0.23\textwidth}
            \centering
            \includegraphics[width=\textwidth]{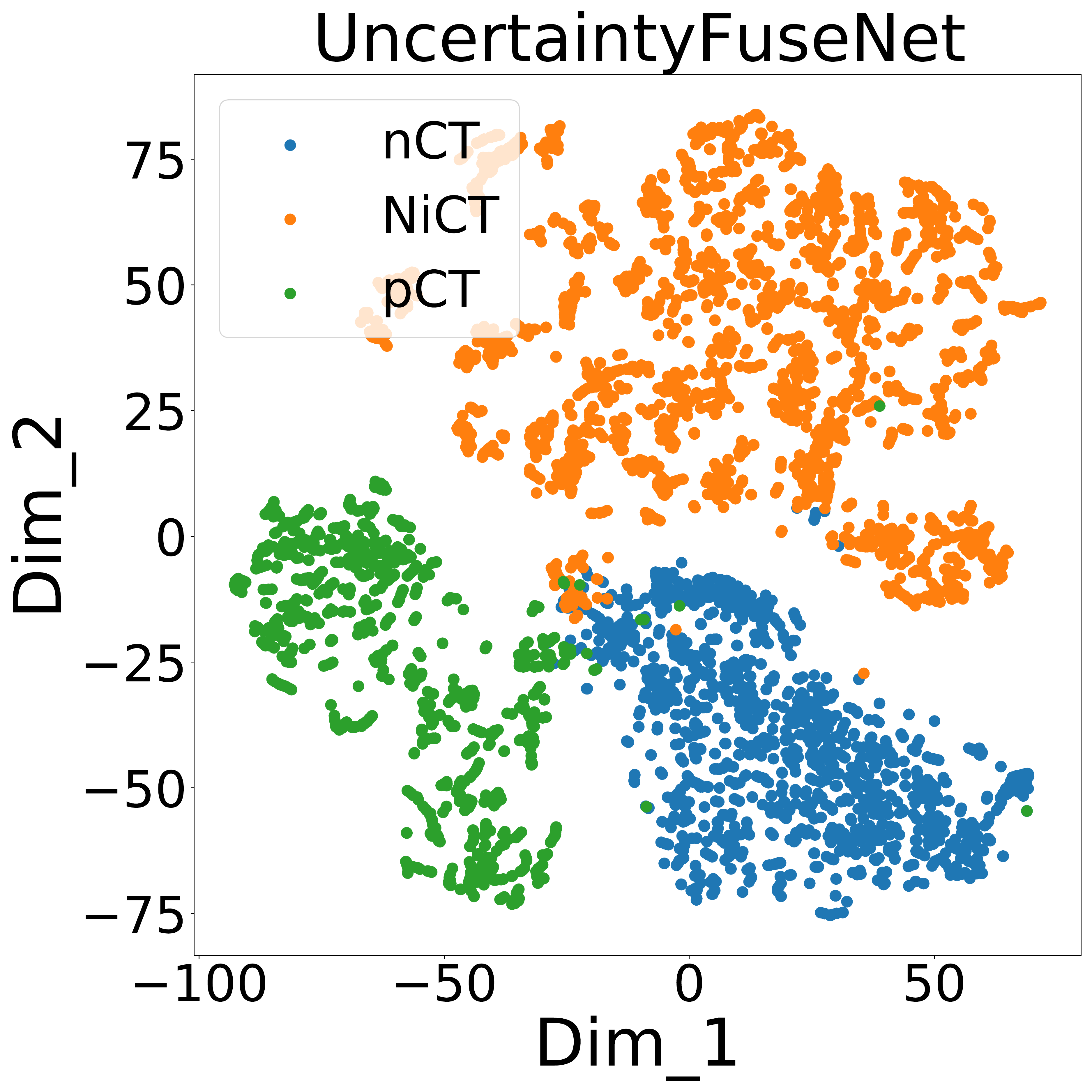}
            \caption{Fusion Model without UQ}
            \label{fig:SSL_F2324}
    \end{subfigure} 
     \begin{subfigure}[b]{0.23\textwidth}
            \centering
            \includegraphics[width=\textwidth]{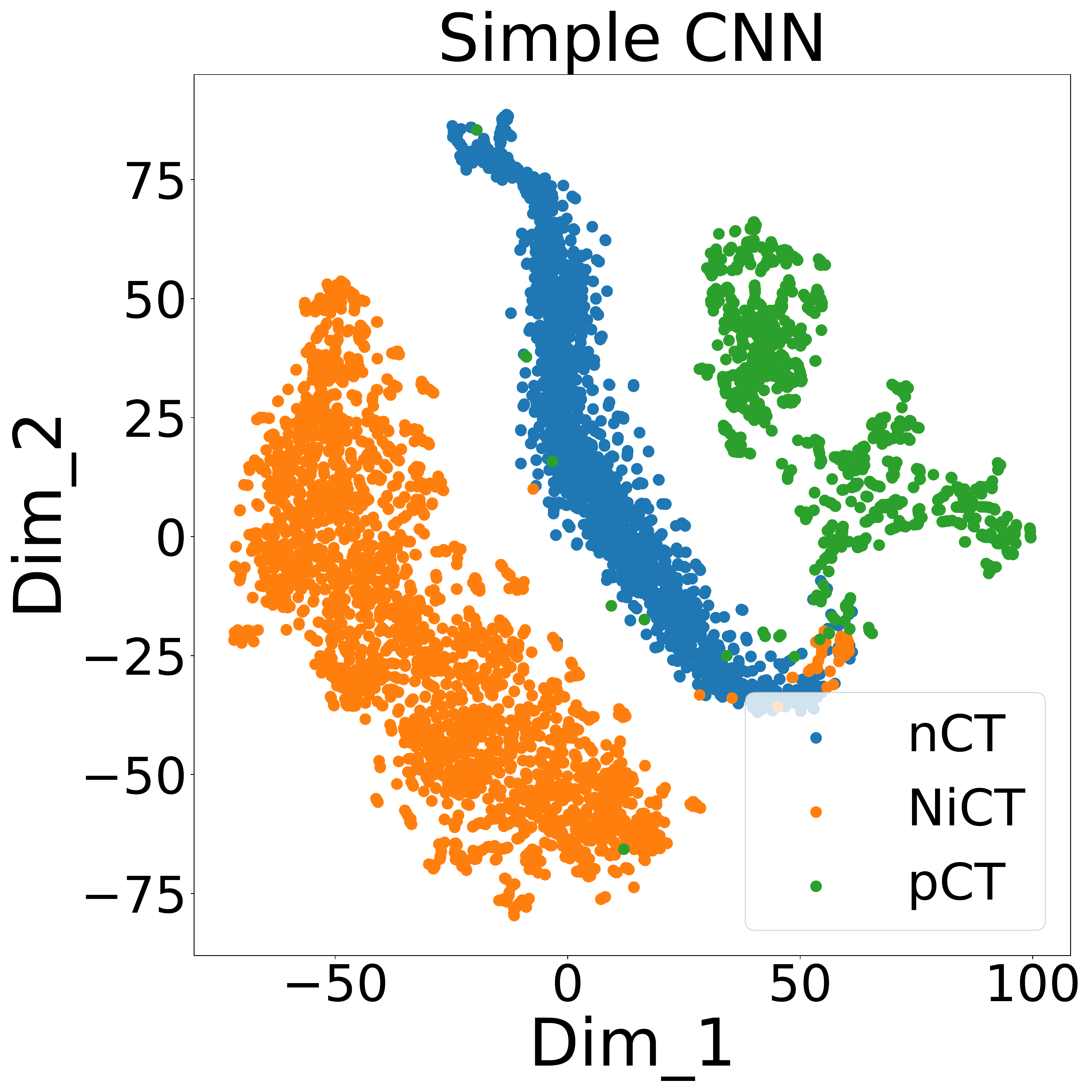}
            \caption{Simple CNN with UQ}
            \label{fig:SSL_F2232}
    \end{subfigure}\\
    \begin{subfigure}[b]{0.23\textwidth}
            \centering
            \includegraphics[width=\textwidth]{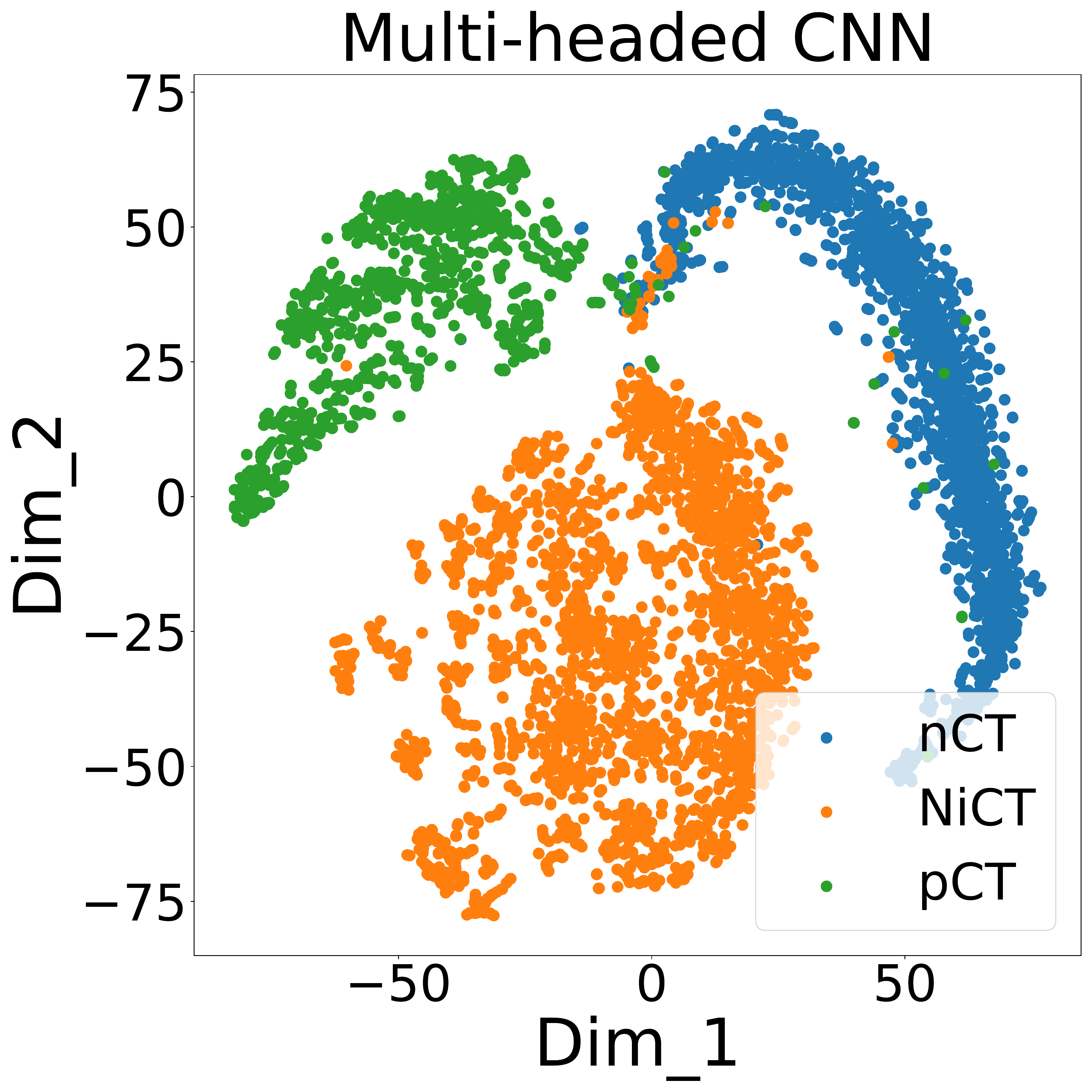}
            \caption{Multi-headed CNN with UQ}
            \label{fig:SSL_F2324}
    \end{subfigure}  
    \begin{subfigure}[b]{0.23\textwidth}
            \centering
            \includegraphics[width=\textwidth]{T-SNE-Visualization-of-Fusion-Model-with-Uncertainty-CT-scan.pdf}
            \caption{Fusion Model with UQ}
            \label{fig:SSL_F2324}
    \end{subfigure}
    \caption{T-SNE visualisation of different models applied to the CT scan data without and with quantifying uncertainty.}\label{fig10}
    \label{COMPA}
\end{figure*}

\begin{figure*}[h!]
\centering
    \begin{subfigure}[b]{0.23\textwidth}
            \centering
            \includegraphics[width=\textwidth]{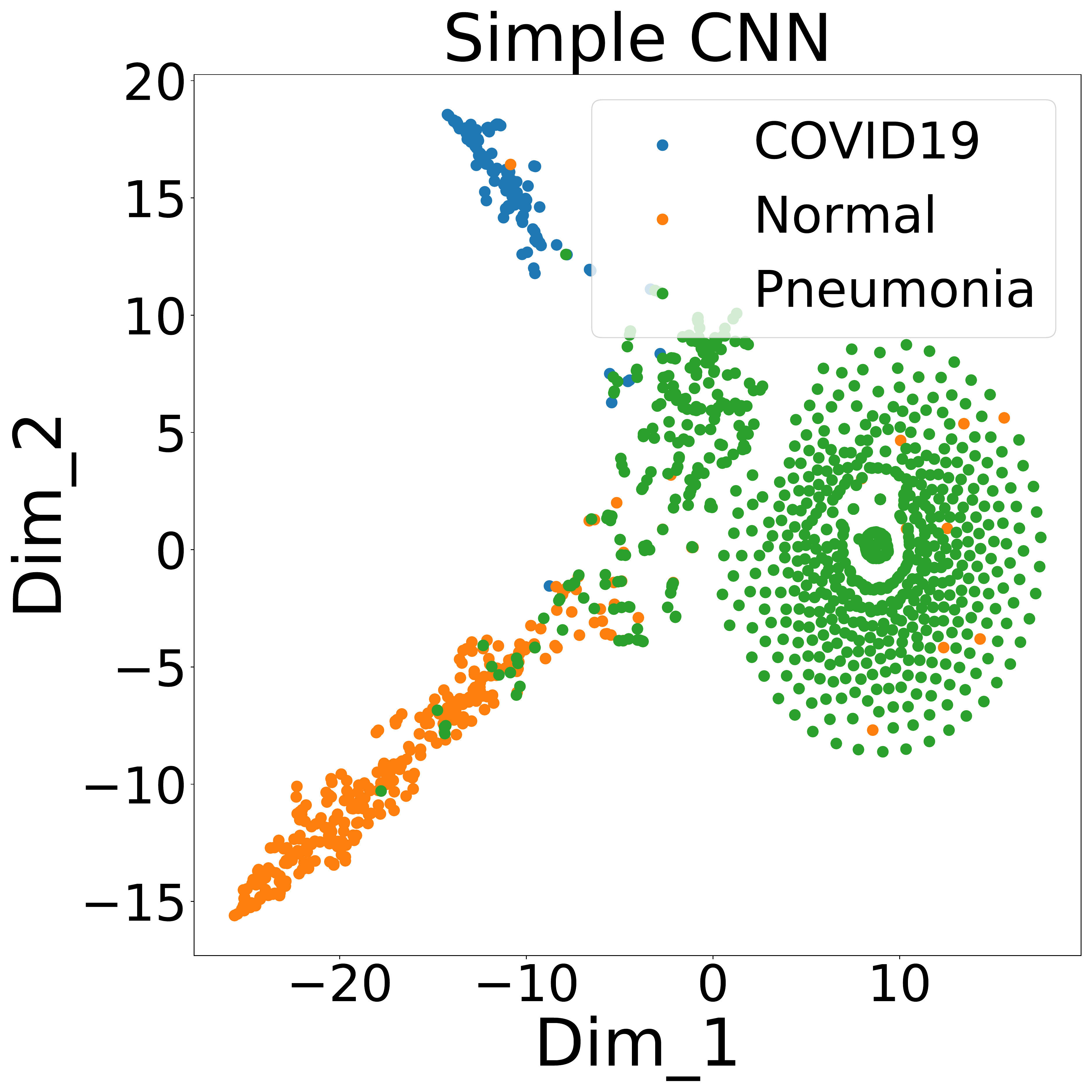}
            \caption{Simple CNN without UQ}
            \label{fig:SSL_F2232}
    \end{subfigure}
    \begin{subfigure}[b]{0.23\textwidth}
            \centering
            \includegraphics[width=\textwidth]{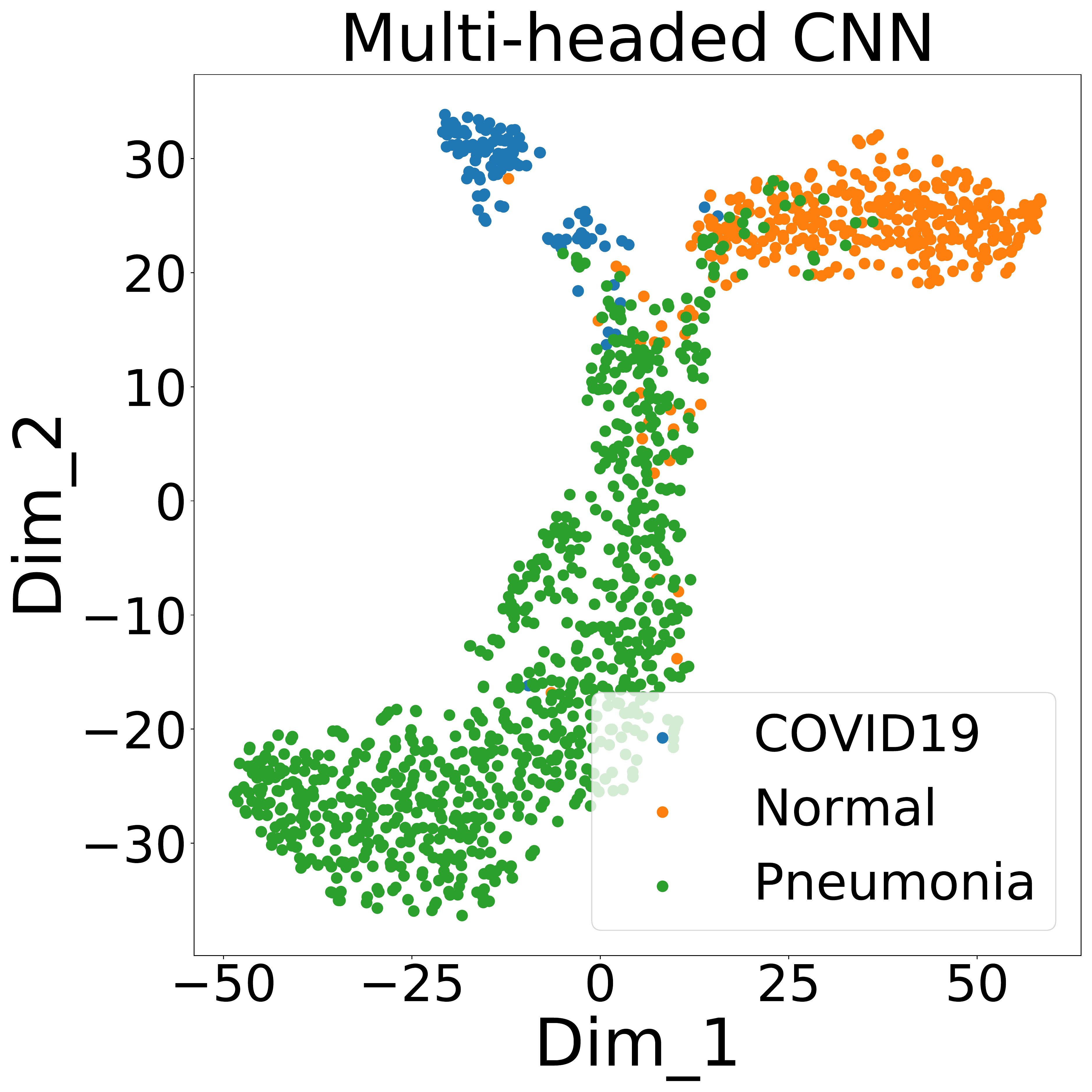}
            \caption{Multi-headed without UQ}
            \label{fig:SSL_F2324}
    \end{subfigure}  
    \begin{subfigure}[b]{0.23\textwidth}
            \centering
            \includegraphics[width=\textwidth]{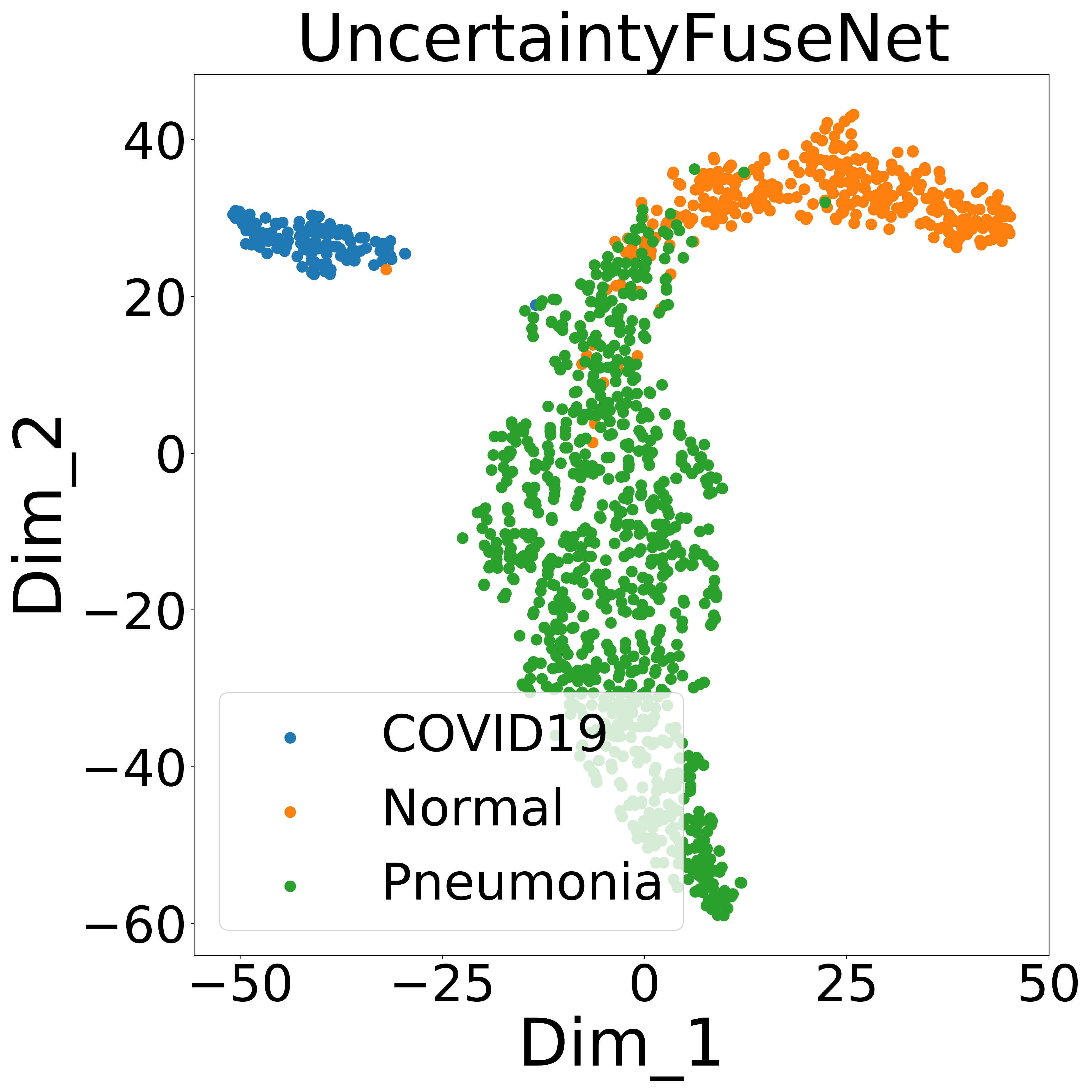}
            \caption{Fusion Model without UQ}
            \label{fig:SSL_F2324} 
    \end{subfigure} 
     \begin{subfigure}[b]{0.23\textwidth}
            \centering
            \includegraphics[width=\textwidth]{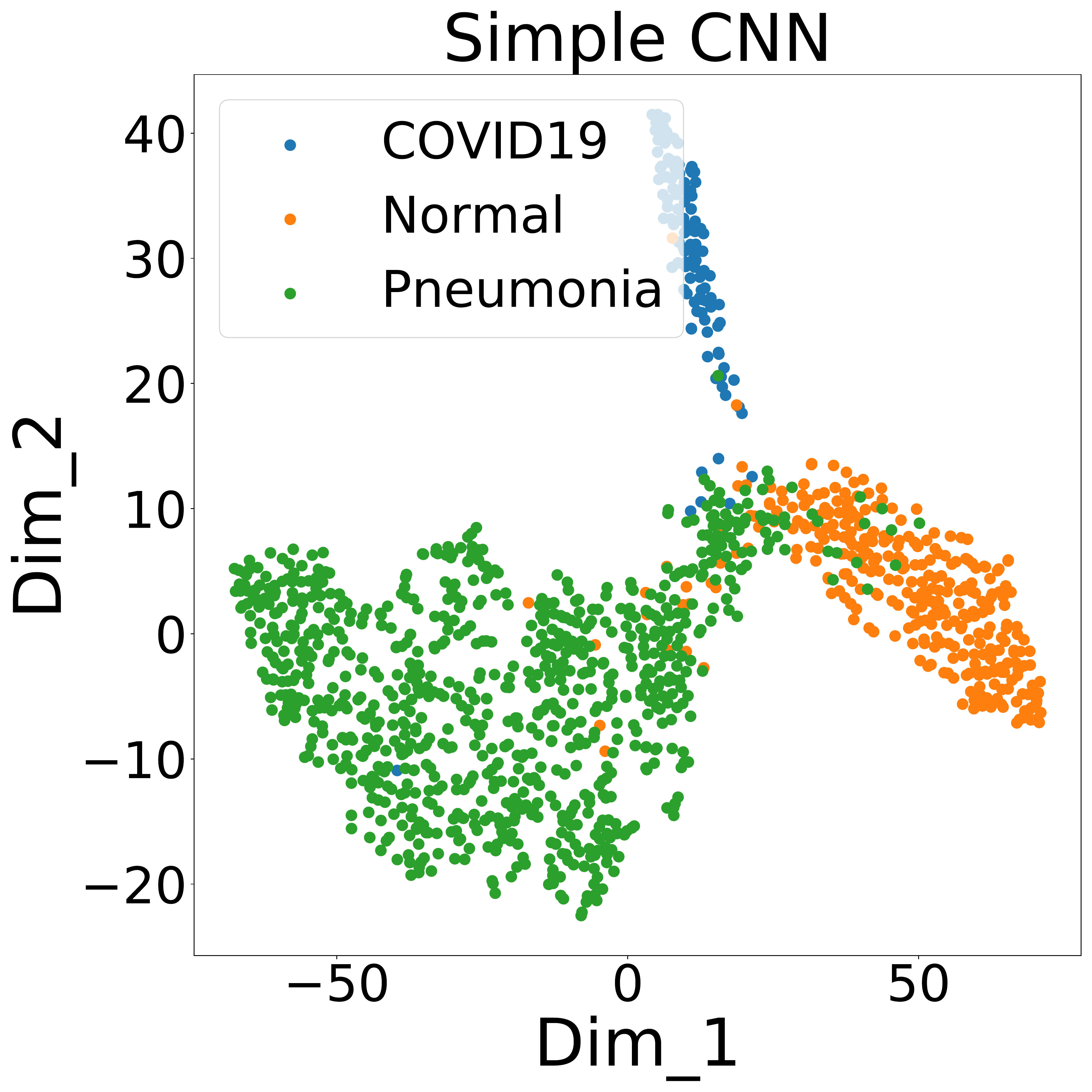}
            \caption{Simple CNN with UQ}
            \label{fig:SSL_F2232}
    \end{subfigure} \\
    \begin{subfigure}[b]{0.23\textwidth}
            \centering
            \includegraphics[width=\textwidth]{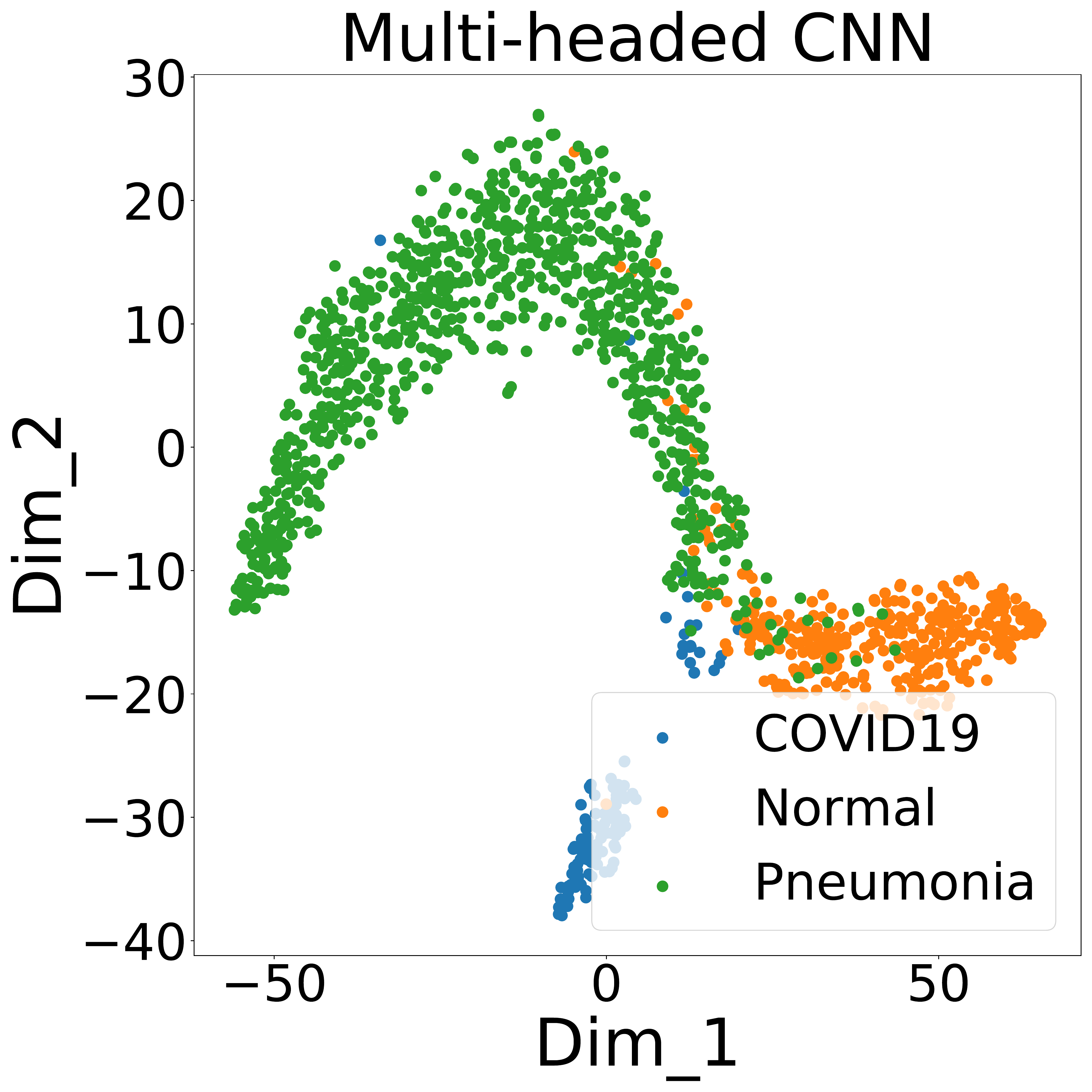}
            \caption{Multi-headed CNN with UQ}
            \label{fig:SSL_F2324}
    \end{subfigure}  
    \begin{subfigure}[b]{0.23\textwidth}
            \centering
            \includegraphics[width=\textwidth]{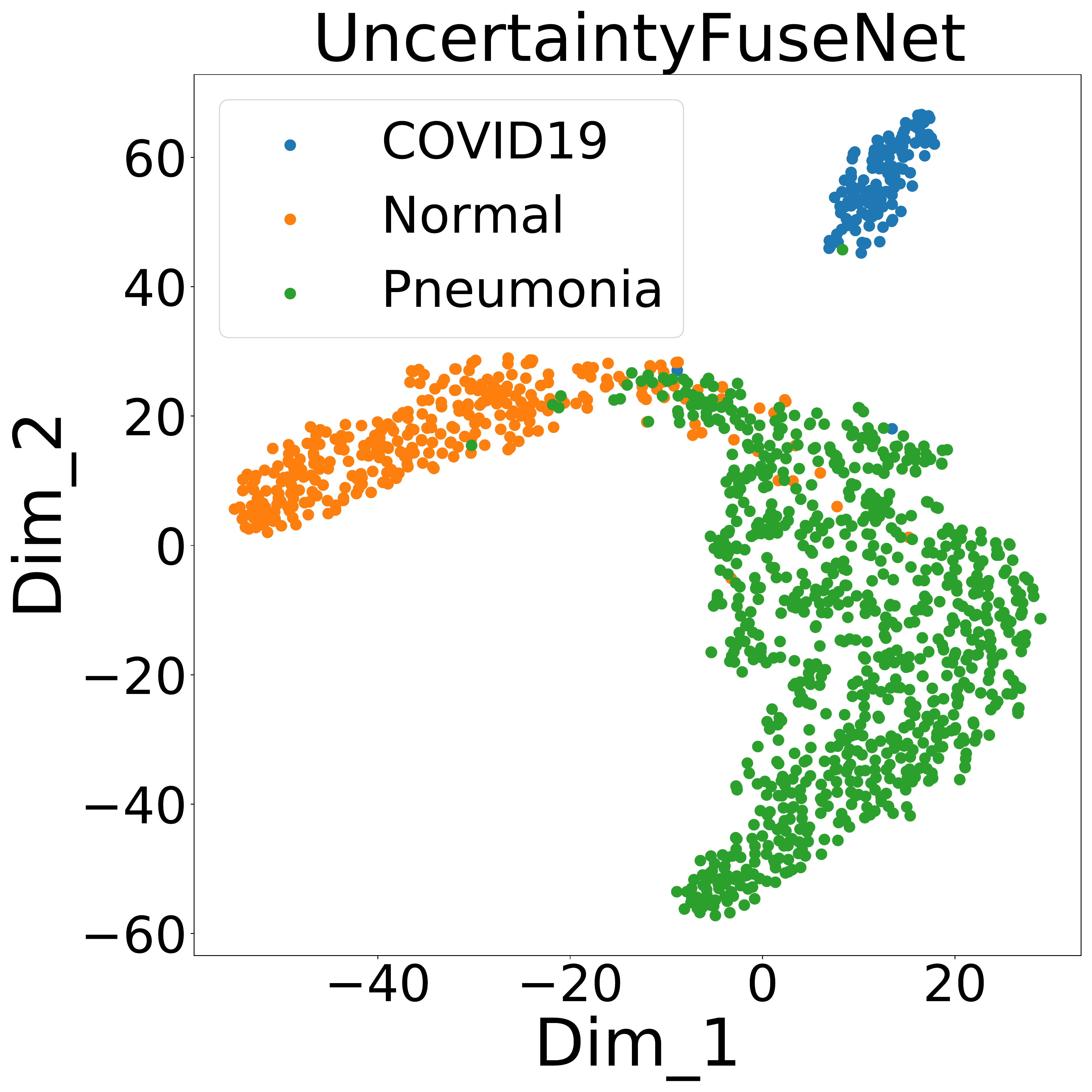}
            \caption{Fusion Model with UQ}
            \label{fig:SSL_F2324}
    \end{subfigure}
    \caption{T-SNE visualisation of different models applied to the X-ray data without and with quantifying uncertainty.}\label{fig11}
\end{figure*}

\begin{figure}[!h]
\centering
    \begin{subfigure}[b]{0.24\textwidth}
            \centering
            \includegraphics[width=\textwidth]{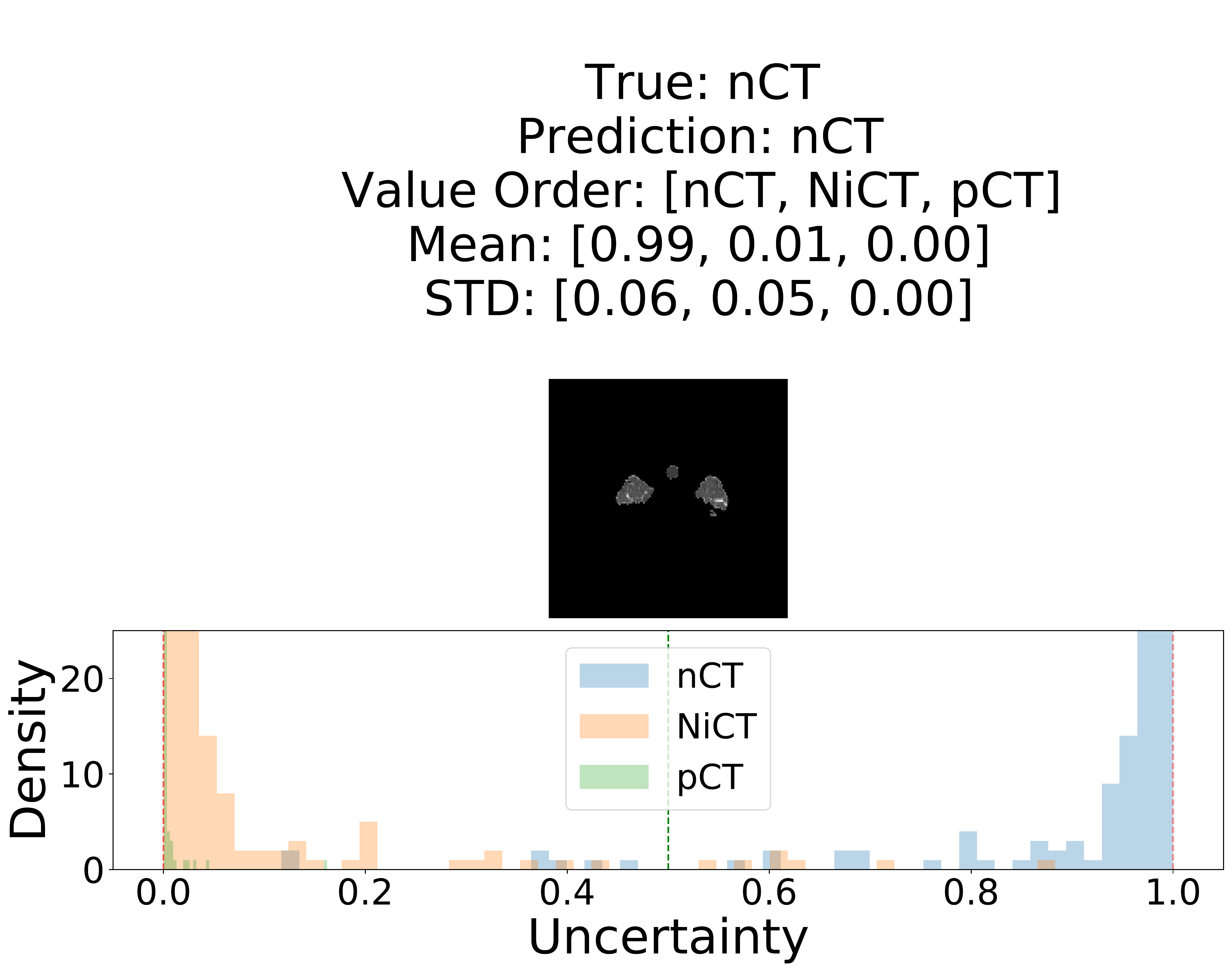}
            \caption{nCT (CT)}
            \label{U1}
    \end{subfigure}
    \begin{subfigure}[b]{0.24\textwidth}
            \centering
            \includegraphics[width=\textwidth]{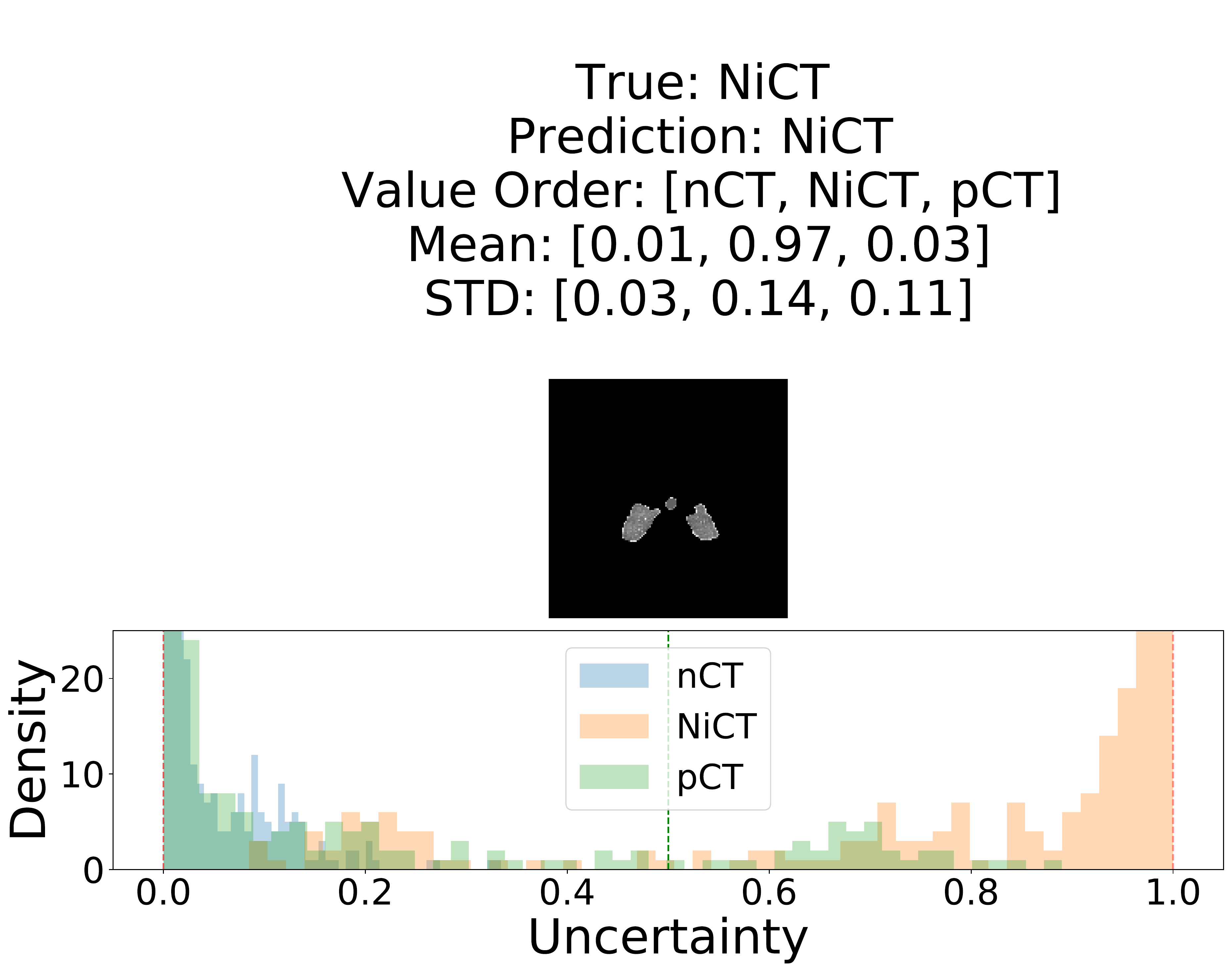}
            \caption{NiCT (CT)}
                \label{U2}
    \end{subfigure}  \\
     \begin{subfigure}[b]{0.24\textwidth}
            \centering
            \includegraphics[width=\textwidth]{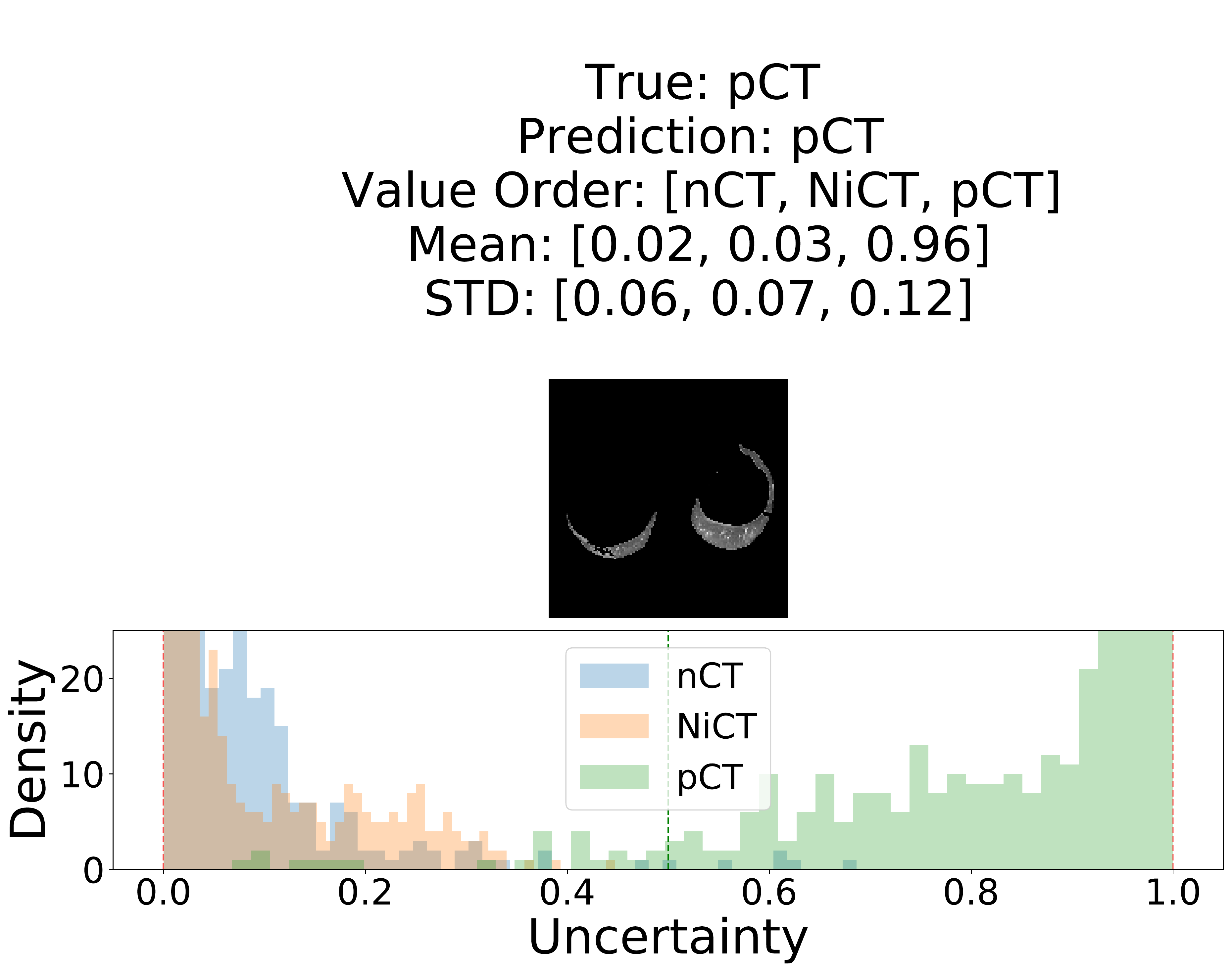}
            \caption{pCT (CT)}
            \label{U3}
    \end{subfigure}
    \begin{subfigure}[b]{0.24\textwidth}
            \centering
            \includegraphics[width=\textwidth]{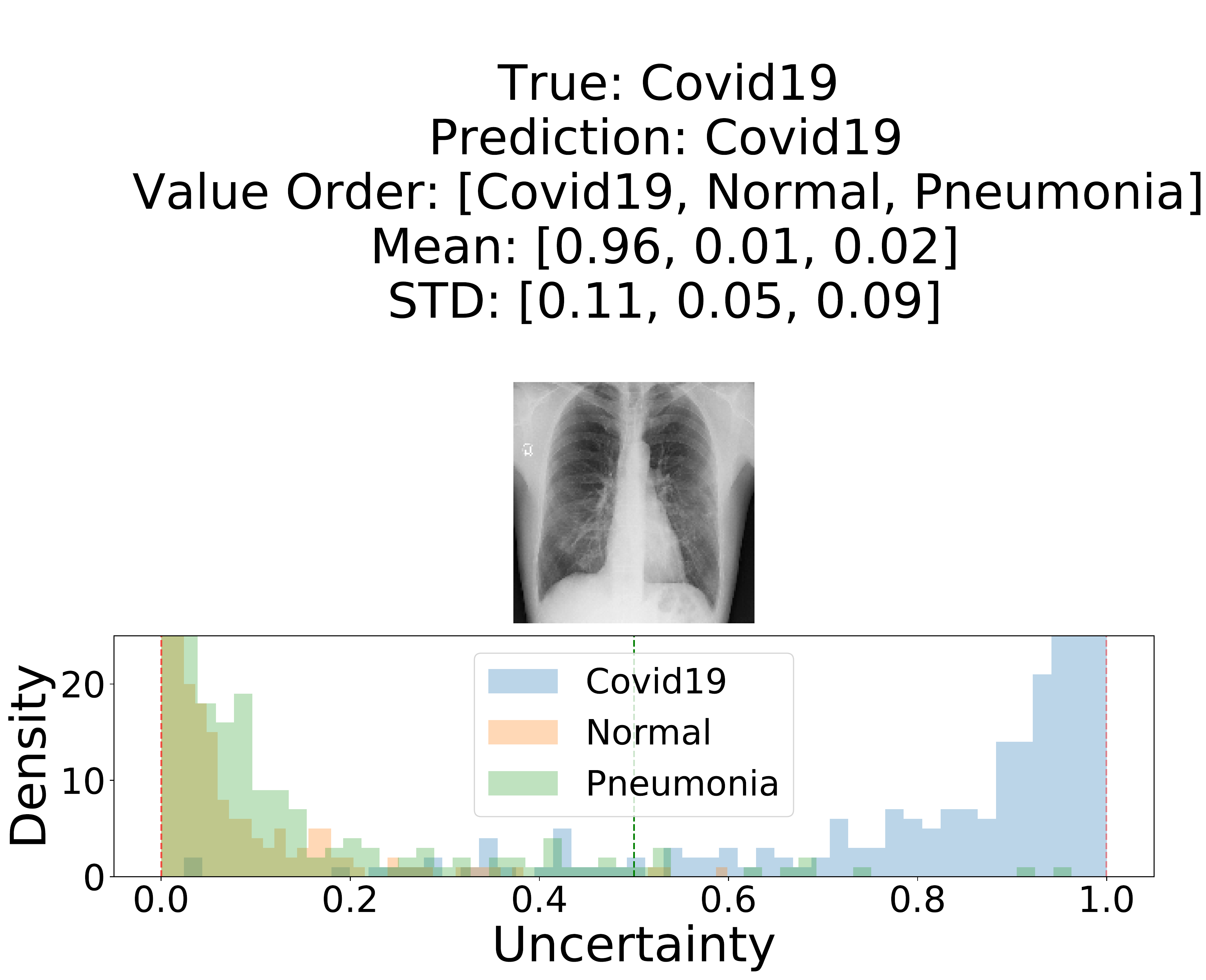}
            \caption{COVID-19 (X-ray)}
            \label{U4}
    \end{subfigure} \\
    \begin{subfigure}[b]{0.24\textwidth}
            \centering
            \includegraphics[width=\textwidth]{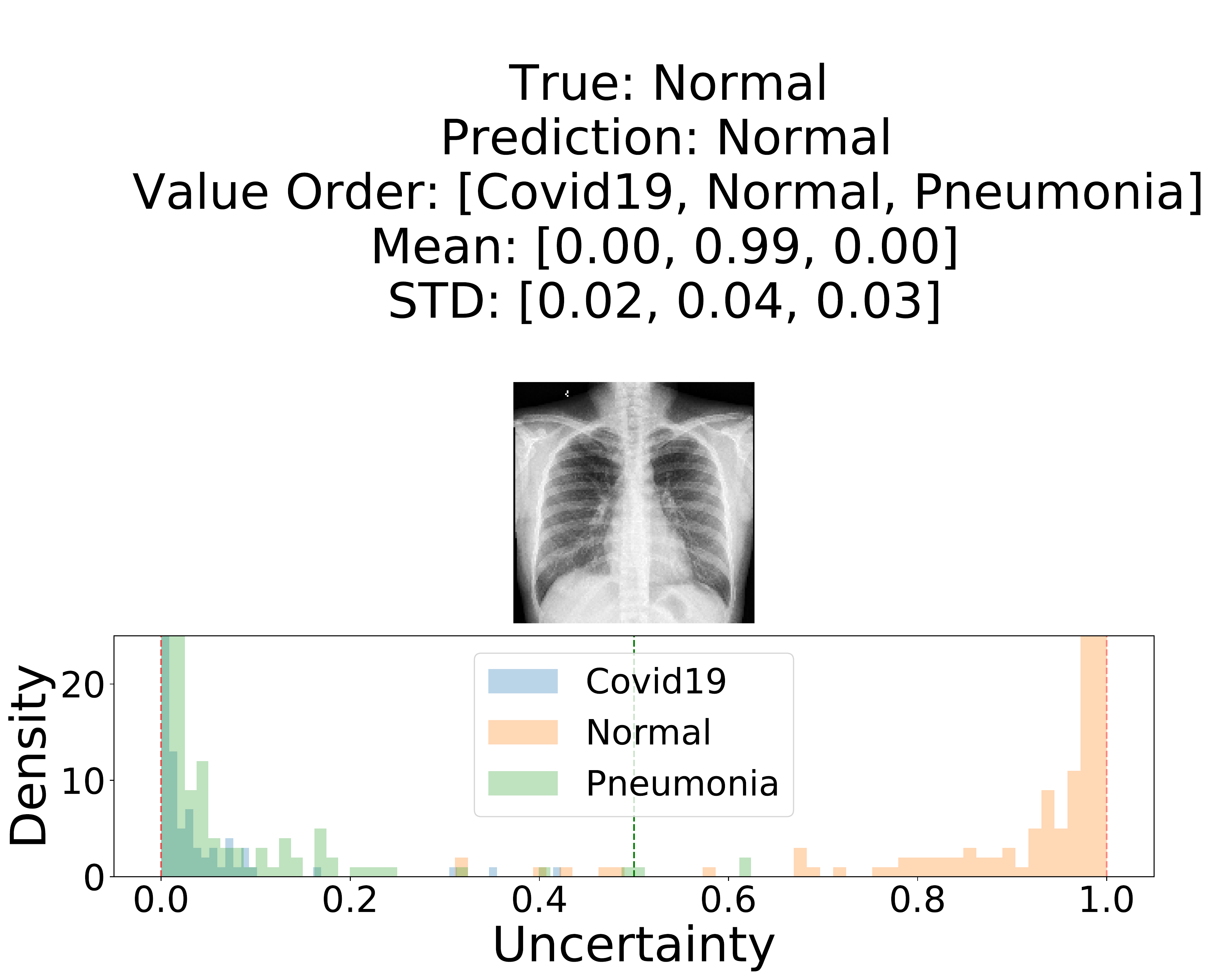}
            \caption{Normal (X-ray)}
            \label{U5}
    \end{subfigure} 
        \begin{subfigure}[b]{0.24\textwidth}
            \centering
            \includegraphics[width=\textwidth]{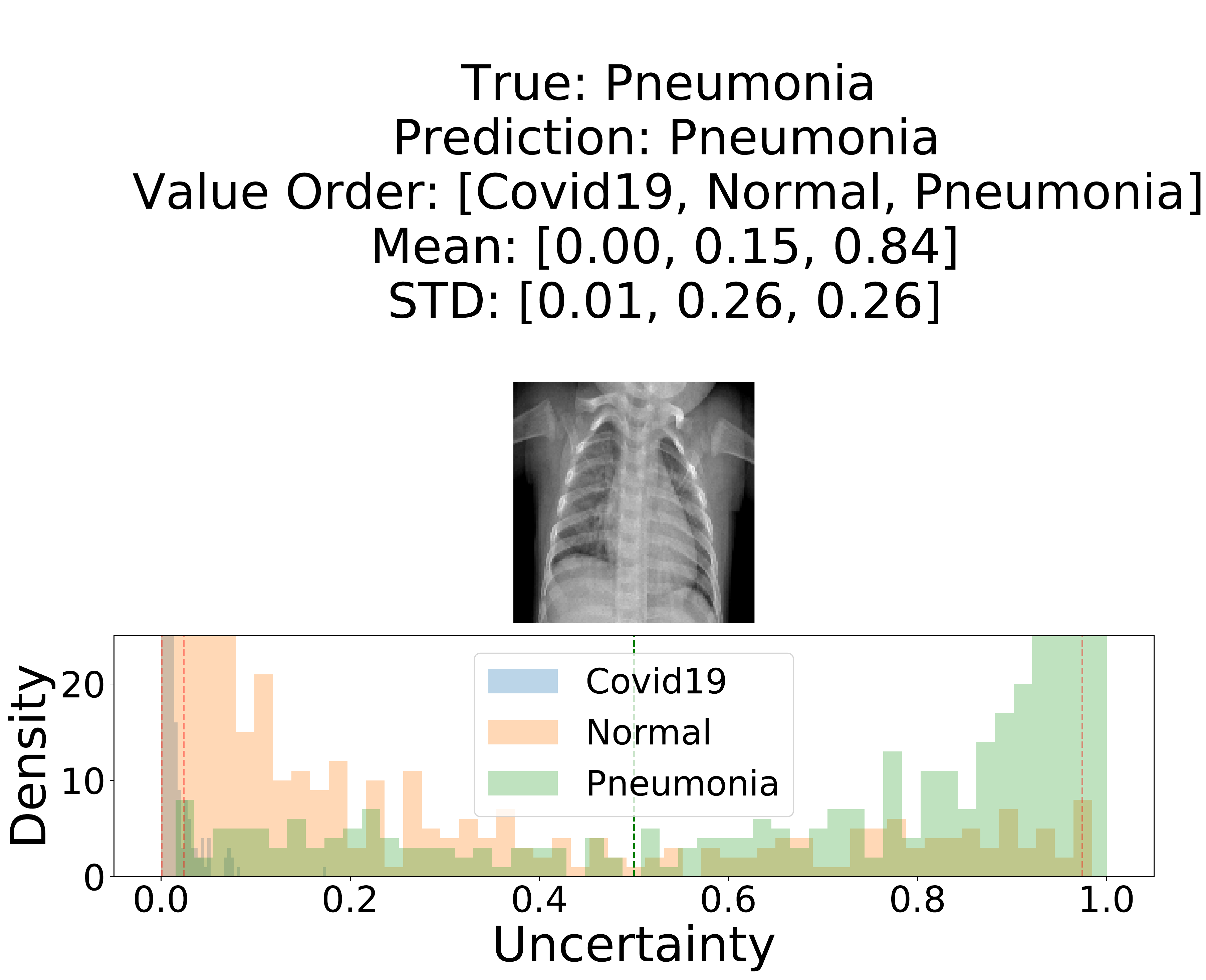}
            \caption{Pneumonia (X-ray)}
            \label{U6}
    \end{subfigure}
    \caption{The output posterior distributions of our proposed feature fusion model calculated for the nCT (\ref{U1}), NiCT (\ref{U2}) and pCT (\ref{U3}) data classes for the CT scan dataset, and the COVID-19 (\ref{U4}), Normal (\ref{U5}) and Pneumonia (\ref{U6}) data classes for the X-ray dataset.}\label{POOUT}
\end{figure}

Nowadays, timely and accurate detection of COVID-19 cases has become a crucial health care task. Various methods from different fields of science have been proposed to tackle the problem of accurate COVID-19 diagnostic. Traditional machine learning (ML) and deep learning (DL) methods have been among the most effective of them. In this work, we mainly focused on the detection of COVID-19 cases using CT scan and X-ray image data. We proposed a new simple but very efficient feature fusion model, called UncertaintyFuseNet, and compared its performance with several classical ML and DL techniques. The prediction results we obtained confirm that our feature fusion model can be highly effective in detecting the COVID-19 cases. Moreover, we have shown the superiority of our model in dealing with noise data. The obtained results also reveal that the proposed UncertaintyFuseNet model can be effectively used for classifying previously unseen images. \\
Our study attempts to fill the gap reported in the literature ~\cite{wang2021covid}. To do so, we have compared the performance of the proposed feature fusion model to recent state-of-the-art machine learning techniques used to classify CT scan and X-ray image data (see Table II in Supplementary Material). Figs. 1 to 6 in the Supplementary Material present confusion matrices for the considered CT scan X-ray datasets computed with and without quantifying uncertainties, and Grad-CAM visualization. The Grad-CAM visualization procedure was carried out to identify the important features for each data class (this analysis was conducted for both CT scan and X-ray image datasets). Figs. 5 and 6 in the Supplementary Material illustrate the most important features used by our feature fusion model to identify each data class separately. Moreover, the T-SNE visualisation of different models applied to the CT scan and X-ray datasets without and with quantifying uncertainty are presented in Figs. \ref{fig10} and \ref{fig11}. Finally, the output posterior distributions of our proposed feature fusion model for both considered image datasets are presented in Fig. \ref{POOUT}. This figure clearly shows that the correctly classified samples of a given class do not overlap with samples of the other classes (incorrect classes). 

\subsection{Comparison with the State-of-the-Art}
In this sub-section, we quickly compare the results provided by our new model with those yielded by the state-of-the-art DL techniques (see Tables \ref{COMRRRRRCT} and \ref{COMRRRRRXRAY}). The state-of-the-art models used in our comparison are the Bayesian Deep Learning \cite{ghoshal2020estimating}, DarkCovidNet \cite{ozturk2020automated},CNN \cite{jadon2021covid}, DeTraC (Decompose, Transfer, and Compose) \cite{abbas2021classification}, and ResNet50 \cite{narin2021automatic} models.

\begin{table}[h!] 
\scriptsize
\caption{Comparison of the results of our DL feature fusion model with the state-of-the-art DL models for CT scan data.}
\begin{tabular}{lllll}
\hline
DL Model & Precison & Recall & F-Measure & Accuracy \\ \hline
Bayesian Deep Learning \cite{ghoshal2020estimating} & 98.351 &  98.333& 98.342 &  98.333  \\
DarkCovidNet \cite{ozturk2020automated} & 97.460 & 97.458& 97.459 & 97.458  \\
CNN \cite{jadon2021covid} & 97.753 & 97.750  & 97.751 & 97.750   \\
DeTraC \cite{abbas2021classification} & 96.972 & 96.958 & 96.965 & 96.958   \\
 ResNet50 \cite{narin2021automatic} &95.571  &  95.541& 95.556 &   95.541 \\
\textbf{Proposed Feature Fusion Model} & \textbf{99.085} & \textbf{99.085} &  \textbf{99.085} & \textbf{99.085}  \\ \hline
\label{COMRRRRRCT}
\end{tabular}
\end{table}

\begin{table}[h!] 
\scriptsize
\caption{Comparison of the results of our DL feature fusion model with the state-of-the-art DL models for X-ray data.}
\begin{tabular}{lllll}
\hline
DL Model & Precison & Recall & F-Measure & Accuracy \\ \hline
Bayesian Deep Learning \cite{ghoshal2020estimating} & 95.398  &95.419  & 95.408 & 95.419   \\
DarkCovidNet \cite{ozturk2020automated} & 95.752 & 95.729 &95.741  & 95.729  \\
CNN \cite{jadon2021covid} &95.400  & 95.341 & 95.370 &  95.341  \\
DeTraC \cite{abbas2021classification} & 95.276 & 95.263 & 95.270 &  95.263  \\
ResNet50 \cite{narin2021automatic} & 94.153 & 94.177 & 94.165 &  94.177  \\
\textbf{Proposed Feature Fusion Model} & \textbf{96.350} & \textbf{96.370} & \textbf{96.360} & \textbf{96.350} \\ \hline
\label{COMRRRRRXRAY}
\end{tabular}
\end{table}

As can be seen from Tables \ref{COMRRRRRCT} and \ref{COMRRRRRXRAY}, our proposed feature fusion model not just only  achieved superior performance but also significantly outperformed the state-of-the-art models applied to the same datasets.  

\subsection{Significance of the UncertaintyFuseNet feature fusion model}
 
Wang et al.~\cite{wang2021covid} proposed a DL feature fusion model for COVID-19 case detection. The model introduced by Wang et al. provided excellent prediction performance for CT scan data considered. However, the authors stated that their model may be much less efficient for other types of medical data such as X-ray images. In another study, Tang et al.~\cite{tang2021edl} proposed an ensemble deep learning model for COVID-19 detection using X-ray image data only. Moreover, most of existing studies focus on COVID-19 case detection without conducting any uncertainty analysis of the model's predictions. Shamsi et al.~\cite{shamsi2021uncertainty} have been among rare authors who considered uncertainty in their study; however, they used very small datasets in their training experiments. \\
In this work, we proposed a novel general feature fusion model which can be effectively used to analyze large CT scan and X-ray datasets (both of these types of images can be processed successively), while quantifying the uncertainty of the model's predictions using the Ensemble MC Dropout (EMCD) technique. It should be noted that the proposed feature fusion model could be easily generalized to classify other complex diseases. Moreover, the model's performance could be further improved by incorporating into its different optimization algorithms such as the Arithmetic optimization algorithm~\cite{abualigah2021arithmetic}, Aquila optimizer~\cite{abualigah2021aquila}, Artificial Immune
System (AIS) algorithm~\cite{abdar2019wart}, Marine Predators algorithm~\cite{sahlol2020covid}, or Cuckoo search optimization algorithm~\cite{yousri2021covid}.

The most important features of our UncertaintyFuseNet feature fusion model are summarized below:
\begin{enumerate} 
  \item Our model provided the highest COVID-19 detection performance compared to traditional machine learning models, some simple deep learning models as well as to state-of-the-art deep learning techniques for both considered types of medical data (CT scan and X-ray images).
    \item Proposed model take advantage of an uncertainty quantification strategy based on the effective Ensemble MC Dropout (EMCD) technique.
    \item Proposed model is robust against noise.
    \item Proposed model is able to detect unknown data with high accuracy.
\end{enumerate}

\section{Conclusion}
\label{Sec:Co}
In this study, we have described a new deep learning feature fusion model to accurately detect COVID-19 cases using CT scan and X-ray data. In order to detect the COVID-19 cases accurately and provide health practitioners with an efficient diagnostic tool they could rely on, we carried out the uncertainty quantification of the model's predictions while detecting the disease cases. Moreover, our UncertaintyFuseNet model demonstrated an excellent robustness to noise and ability to process unknown data. A class-wise analysis procedure has been implemented to ensure a steady performance of the model. We have demonstrated the effectiveness of our model using various computational experiments. Our experimental results suggest that the presented feature fusion model can be applied to analyze efficiently both CT and X-ray data. The use of hierarchical features in the model's architecture helped UncertaintyFuseNet to outperform the considered traditional machine learning models, classical deep learning models, and state-of-the-art deep learning models. The limitations of the proposed feature fusion model will be addressed in our future studies. Thus, in the future, we intend to: (i) expand the considered COVID-19 datasets and test our feature fusion model using multi-modal data, (ii) include an attention mechanism while merging features, and (iii) integrate into our model some modern data fusion techniques such as decision level fusion. 

\section{Acknowledgments}
This research was supported by the Australian Research Council’s Discovery Projects funding scheme (project DP190102181).

%





\ifCLASSOPTIONcaptionsoff
  \newpage
\fi

\bibliographystyle{IEEEtran}
\bibliography{ref.bib}

\begin{thebibliography}{10}
\bibitem{shanshanoriginal}
Z.~Shanshan, ``Original paper multi-source information fusion technology and
  its engineering application.''
\end{thebibliography}

\end{document}